\documentclass[journal]{IEEEtran}
\usepackage{cite}
\usepackage[pdftex]{graphicx}
\usepackage[cmex10]{mathtools}
\usepackage{amssymb}
\usepackage{epsfig}
\usepackage{subfigure}
\usepackage{algorithm}
\usepackage{algpseudocode}
\usepackage[usenames,dvipsnames]{color}
\usepackage{array}

\usepackage{bm}
\usepackage{bbm}
\usepackage{microtype}
\usepackage{kotex}
\usepackage[T1]{fontenc}
\usepackage[utf8]{inputenc}

\newtheorem{lemma}{Lemma}
\newtheorem{prop}{Proposition}

 \def\cN{{\mathcal{N}}}

 \def\bb{{\mathbf{b}}} \def\bc{{\mathbf{c}}} \def\bd{{\mathbf{d}}}
\def\bee{{\mathbf{e}}} \def\bff{{\mathbf{f}}} \def\bg{{\mathbf{g}}} \def\bh{{\mathbf{h}}}
   
 \def\bn{{\mathbf{n}}}  
 \def\br{{\mathbf{r}}}  \def\bt{{\mathbf{t}}}
\def\bu{{\mathbf{u}}} \def\bv{{\mathbf{v}}} \def\bw{{\mathbf{w}}} 
 \def\bz{{\mathbf{z}}}  

\def\bA{{\mathbf{A}}} \def\bB{{\mathbf{B}}} \def\bC{{\mathbf{C}}} \def\bD{{\mathbf{D}}}
 \def\bF{{\mathbf{F}}}  
\def\bI{{\mathbf{I}}}  \def\bK{{\mathbf{K}}} 
   
\def\bQ{{\mathbf{Q}}}   
   
 \def\bZ{{\mathbf{Z}}}

\DeclareMathOperator*{\argmin}{arg\,min}
\DeclareMathOperator*{\argmax}{arg\,max}

\begin{document}


\title{Adaptive Beam Design for V2I Communications {using} {Vehicle Tracking} with Extended Kalman Filter}


\author{Seong-Hwan Hyun,~\IEEEmembership{Student Member,~IEEE,} Jiho Song,~\IEEEmembership{Member,~IEEE,} Keunwoo Kim,~\IEEEmembership{Student Member,~IEEE,}
\\
Jong-Ho Lee,~\IEEEmembership{Member,~IEEE,} and Seong-Cheol Kim,~\IEEEmembership{Senior Member,~IEEE}
\thanks{Copyright (c) 2015 IEEE. Personal use of this material is permitted. However, permission to use this material for any other purposes must be obtained from the IEEE by sending a request to pubs-permissions@ieee.org.}
\thanks{This work has been submitted to the IEEE for possible publication. Copyright may be transferred without notice, after which this version may no longer be accessible.}
\thanks{S.-H.\ Hyun,  K.  \ Kim, and S.-C. \ Kim are with the School of Electrical Engineering, Seoul National University, and also with the Institute of New Media \& Communications
(INMC), Seoul 08826, South Korea (e-mail: \{shhyun, kimkeunwoo15,  sckim\}@maxwell.snu.ac.kr).}
\thanks{J.\ Song \textit{(corresponding author)} is with the School of Electrical Engineering, University of Ulsan, Ulsan 44610, South Korea (e-mail: jihosong@ulsan.ac.kr).}
\thanks{J.-H.\ Lee  is with the School of Electronic Engineering, Soongsil University, Seoul 06978, South Korea (e-mail: jongho.lee@ssu.ac.kr).}}

\maketitle


\begin{abstract}
Vehicle-to-everything communication system is a strong candidate for improving the driving experience and automotive safety by linking vehicles to wireless networks.
To take advantage of the full benefits of vehicle connectivity, it is essential to ensure a stable network connection between roadside unit (RSU) and fast-moving vehicles.
Based on the extended Kalman filter (EKF), we develop a vehicle tracking algorithm to enable reliable radio connections.
For the vehicle tracking algorithm, we focus on estimating the rapid changes in the beam direction of a high-mobility vehicle while reducing the feedback overhead.
Furthermore, we design a beamforming codebook that considers the road layout and RSU.
By leveraging the proposed beamforming codebook, vehicles on the road can expect a service quality similar to that of conventional cellular services.
Finally, a beamformer selection algorithm is developed to secure sufficient gain for the system's link budget.
Numerical results verify that the EKF-based vehicle tracking algorithm and the proposed beamforming structure are more suitable for vehicle-to-infrastructure networks compared to existing schemes.
\end{abstract}


\IEEEpeerreviewmaketitle

\section{Introduction}

Intelligent transportation system (ITS) has been considered an emerging technology that can improve the driving experience \cite{Ref_Lu14,Ref_R2,Ref_Zhang11}.
The objective of ITS is to enhance traffic efficiency and road safety by collecting sensor data associated with traffic conditions \cite{Ref_Rauch11}.
While the sensor data increases situational awareness of each vehicle, sharing the collected data between neighboring vehicles and/or roadside unit (RSU) can help facilitate fully automated driving.
It is also necessary to obtain traffic information beyond the sensor range to build networks for connected vehicles.
For these reasons, vehicle-to-everything (V2X) communication system is a key technology for supporting ITS, because it enables vehicles to be aware of their surroundings via wireless links \cite{Ref_Chg16Mag,Ref_TS_22_186,Ref_TS_22_185,Ref_TS_22_885}.

To ensure robust communications in V2X systems, it is essential to reevaluate wireless networks \cite{Ref_Schulz17}.
For road safety, low-latency and high-reliability are required to rapidly and accurately transmit warning messages about hazardous situations such as collisions.
The challenge in V2X communication systems is to meet these conflicting service requirements \cite{Ref_Pocovi18}.
For example, obtaining high-resolution channel state information would help maintain a reliable network connection, but it also increases latency, which restricts the use of wireless resources, and vice versa.

Extensive data should be exchanged between RSU and vehicles to provide self-driving and platoon-based driving services \cite{Ref_Chg16Mag,Ref_Lee20}.
Using millimeter{-}wave (mmWave) systems with wide bandwidths is a lucrative solution for offering higher-than-expected data-rate in V2X communications.
However, there are some obstacles to be addressed before providing vehicular communication services at mmWave frequencies \cite{Ref_Ant17}.
Compared to conventional wireless networks operated in the ultra-high frequency (UHF) band, the mmWave band is more sensitive to propagation conditions such as foliage, rainfall, and moisture in the air \cite{Ref_Zhang15,Ref_Rappaport15}.
Due to the short wavelength, the antenna cross-section becomes smaller, resulting in a larger path loss than with UHF signals.
To compensate for the excessive propagation loss, it is necessary to consider a multi-antenna system for directive gain \cite{Ref_Son14}.

A very narrow beam width for directional transmission enhances the beamforming gain, but may lead to intermittent radio link disconnections \cite{Ref_Va17}.
The high-mobility in V2X communications necessitates more beam training samples to reliably track vehicle movement.
However, this beam alignment significantly increases the beam training overhead.
In order to resolve this trade-off problem, a vehicle tracking algorithm with a low overhead is required to make V2X communication systems robust to fast-moving vehicle environments \cite{Ref_Jayaprakasam17,Ref_Va16}.

Vehicle tracking algorithms for mmWave broadband systems have been developed based on the extended Kalman filter (EKF)  \cite{Ref_Va16,Ref_Zha16,Ref_Sha19}.
Although previous works in \cite{Ref_Va16,Ref_Zha16} are eligible for estimating beam directions with moderate computational complexity, the {vehicle tracking} frameworks seem inappropriate because the dynamic state evolution model is designed as a linear function of  beam directions \cite{Ref_Sha19}.
The  vehicle tracking algorithm in \cite{Ref_Sha19} is developed based on a realistic state evolution model, which describes the vehicle motion states, such as position and velocity.
However, the vehicle tracking algorithm  can only estimate the beam direction in the vertical domain because the system model is developed based on a simplified (one-dimensional) geographic channel model.
In practical vehicle-to-infrastructure (V2I) communication scenarios, it is even more important to estimate the beam direction in the horizontal domain because the transmit RSUs are deployed along the travel direction of vehicles.
In this paper, we exploit spatial information for V2I communications to study a practical vehicle tracking framework.
With leveraging spatial information, we propose a vehicle tracking algorithm that can facilitate directional transmission by considering the beam directions in the vertical and horizontal domains.

EKF-based vehicle tracking algorithms are developed based on the linear approximation of the estimated channel vector \cite{Ref_Va16,Ref_Zha16,Ref_Sha19}.
Therefore, the vehicle tracking performance is highly dependent on the current beam estimate \cite{Ref_Kay93,Ref_Ter08}.
Considering the sparse characteristics of mmWave channels, the transmit beamformer for the state update process should be carefully designed to maintain reliable vehicle tracking performance \cite{Ref_Cho14,Ref_Lar19}.
The transmit beamformer in \cite{Ref_Sha19} is chosen from an array manifold set based on the current beam estimate.
Since the beamformers are limited to the array manifold set, it is difficult to design the sounding signal according to the channel conditions.
Specifically, in mmWave systems with multiple antennas, the beam width of the array response vector is extremely narrow.
However, in a low signal-to-noise ratio (SNR) regime, the array response vector with a narrow beam width intermittently disconnects a link for channel sounding because the spatial selectivity becomes more severe in mmWave systems.
To improve the quality of sounding samples, we develop an adaptive channel sounding algorithm that can design the beam pattern of the beamformer according to the state prediction performance.

The conventional beamforming codebook is designed to achieve uniformly distributed beam directions considering the characteristics of the circular shape cellular service region \cite{Ref_Xiao16,Ref_Xiao17,Ref_Noh17}.
However, in V2I communication scenarios, RSU is deployed near the road; the distance between RSU and the vehicle is considerably shorter than the road length.
Due to the unusual shape of the service region, it is not reasonable to design a beamforming codebook under the assumptions that the beam directions should be uniformly distributed and the beamformers have an equal beam width.
In this paper, the beamformers are designed to have different beam widths by reflecting geographical features to provide a similar level of quality-of-service, regardless of the vehicle location on the road.

The objective of this work is to facilitate a large beamforming gain by designing beamforming codebooks that take into account geographical characteristics as well as selecting the most appropriate beamformer based on the vehicle position obtained through the EKF algorithm utilizing spatial information.
The contributions of our paper are summarized as follows.
\begin{itemize}
\item \textit{{Vehicle tracking} algorithm:}
Since the relative velocity between a high-speed vehicle and RSU is significant, the beam direction towards the vehicle changes rapidly.
RSU needs to know the vehicle motion state in real time to transmit a downlink signal in the right direction.
The changes in the V2I channel  direction can be updated based on a feedback-based vehicle tracking algorithm.
However, it would impose a burden on the feedback link.
We develop a vehicle tracking algorithm that predicts a vehicle's motion state under high-mobility scenarios, using uplink sounding signals with a small amount of vehicle motion feedback to RSU.
In the proposed algorithm, an optimal combiner at RSU is computed to maximize the vehicle tracking performance through the uplink channel sounding process. Moreover, the acceleration parameter estimation can further reduce the estimation uncertainty of position and velocity information.
\item \textit{Beamformer design algorithm:}
The wireless service region in the V2I network is significantly different from those in conventional cellular networks.
Since the majority of beamforming codebooks are designed to cover a circular coverage area, the beam patterns are severely distorted when projected onto the road.
Assuming beam width of beamformers are the same, the projected coverage in the geographical domain varies depending on the vehicle location (e.g., the edge of the road and the center of the road).
To resolve this problem, we develop a codebook design algorithm tailored for V2I communications, with road layout information that is not considered in the conventional beamforming codebook.
\item \textit{Beamformer selection algorithm:}
Frequent beam switching can impose a burden on wireless communication systems with vehicular channels that vary rapidly with vehicle movement.
It is not feasible to frequently recalculate the transmit beamformer.
Considering that the beam switching speed at RSU is slower than that of a fast-moving vehicle, {the transmit beamformer should be selected by predicting the vehicle positions after a few steps.
To maximize the average beamforming gain,} we develop a method to select the best transmit beamformer by predicting the future state vector of the vehicle based on the EKF algorithm.
\end{itemize}

The paper is organized as follows.
In Section \ref{sec:SM}, we present our system model.
Based on the EKF algorithm, Section \ref{sec:beam_tracking_algorithm} proposes a vehicle tracking algorithm for V2I communication systems.
In Section \ref{sec:adaptive_BP},  we propose a transmit beamformer design algorithm  to maximize the beam alignment and data-rate performance.
We evaluate the proposed V2I communication system in Section \ref{sec:numerical}.
Finally,  Section \ref{sec:con} details our conclusions.

\textit{Notation:} $\mathbb{C}$ {is} the field of complex numbers, $\mathbb{R}$ the field of real numbers,  $\mathcal{N}(m,\sigma^2)$ the normal distribution with mean $m$ and variance $\sigma^2$, $\mathrm{U}(a,b)$ the uniform distribution in the range $[a,b]$, $\mathrm{E}[\cdot]$ the expectation operator, {$[a,b)$ the left-closed, right-open interval between $a$ and $b$, $\mathrm{u}(\cdot)$ the unit-step function, $| \cdot |$ the length of a range}, $\mathfrak{v}\{\cdot\}$ the  principal eigenvector of the matrix, $\mathrm{Tr}\{\cdot\}$ the  trace of the matrix, $\mathbf{1}_{a}$ the $a \times 1$ all ones column vector, $\mathbf{0}_{a}$ the $a \times 1$ all zeros column vector, $\mathbf{0}_{a \times b}$ the $a \times b$ all zeros matrix, $\mathbf{I}_{M}$ the $M \times M$ identity matrix, and $\| \cdot \|_p$ the $p$-norm, respectively. Also,  $\bA^{-1}$, $\bA^H$, $\bA^T$,  and $\bA_{a,b}$ denote inverse, conjugate transpose,  transpose, and $(a,b)$-th entry of the matrix $\bA$, respectively.

\section{System Model}
\label{sec:SM}

We consider a V2I wireless network in which RSU has $M$  antennas, and a single  antenna is mounted on the roof of each vehicle.
The input-output expression for data transmission  is defined by
\begin{align}
\label{eq:input_output}
r_{\ell}^{\mathrm{down}}=\sqrt{\rho_{\ell}} \bh_{\ell}^H\bc_{\ell} s +  n_{\ell}^{\mathrm{down}},
\end{align}
where $r_{\ell}^{\mathrm{down}} \in \mathbb{C}$ is the received signal, $\bh_{\ell} \in \mathbb{C}^{M}$ is the downlink multiple-input single-output channel vector,  $\bc_{\ell} \in \mathbb{C}^M$ is the unit-norm  beamforming vector, $s$  is the data symbol that is constrained by $\mathrm{E}[|s|^2]=1$ and $n_{\ell}^{\mathrm{down}} \sim \mathcal{CN}(0,1)$ is the additive white Gaussian noise.
By assuming unit antenna gain at both the transmitter and the receiver, the average SNR can be defined by $\rho_{\ell} \doteq \frac{\varrho}{\sigma_{\mathrm{n}}^2} \big(\frac{\lambda}{4\pi {d}_{\ell}}\big)^n$, where $\varrho$, $\sigma_{\mathrm{n}}^2$, $\lambda$, ${d}_{\ell}$ and $n$ denote the transmit power, noise power, wavelength of the radio signals, distance between the vehicle and RSU, and the path-loss exponent, respectively.

We consider a hybrid beamforming system  in which $N$ radio frequency chains are connected to a baseband beamforming unit \cite{Ref_Alk14,Ref_Aya14}.
The radio frequency chain generates a unit-norm analog beam steering vector, $\bff_{n} \in \mathbb{C}^M$, consisting of $M$ equal gain complex numbers.
The beamforming vector is constructed by combining $N$ analog beam steering vectors  with {a} unit-norm digital weight vector, $\bv \in \mathbb{C}^N$, such that $\bc_{\ell}=[\bff_{1},\cdots, \bff_{N}]\bv$.

\begin{figure}
\centering
\subfigure{\includegraphics[width=0.375\textwidth]{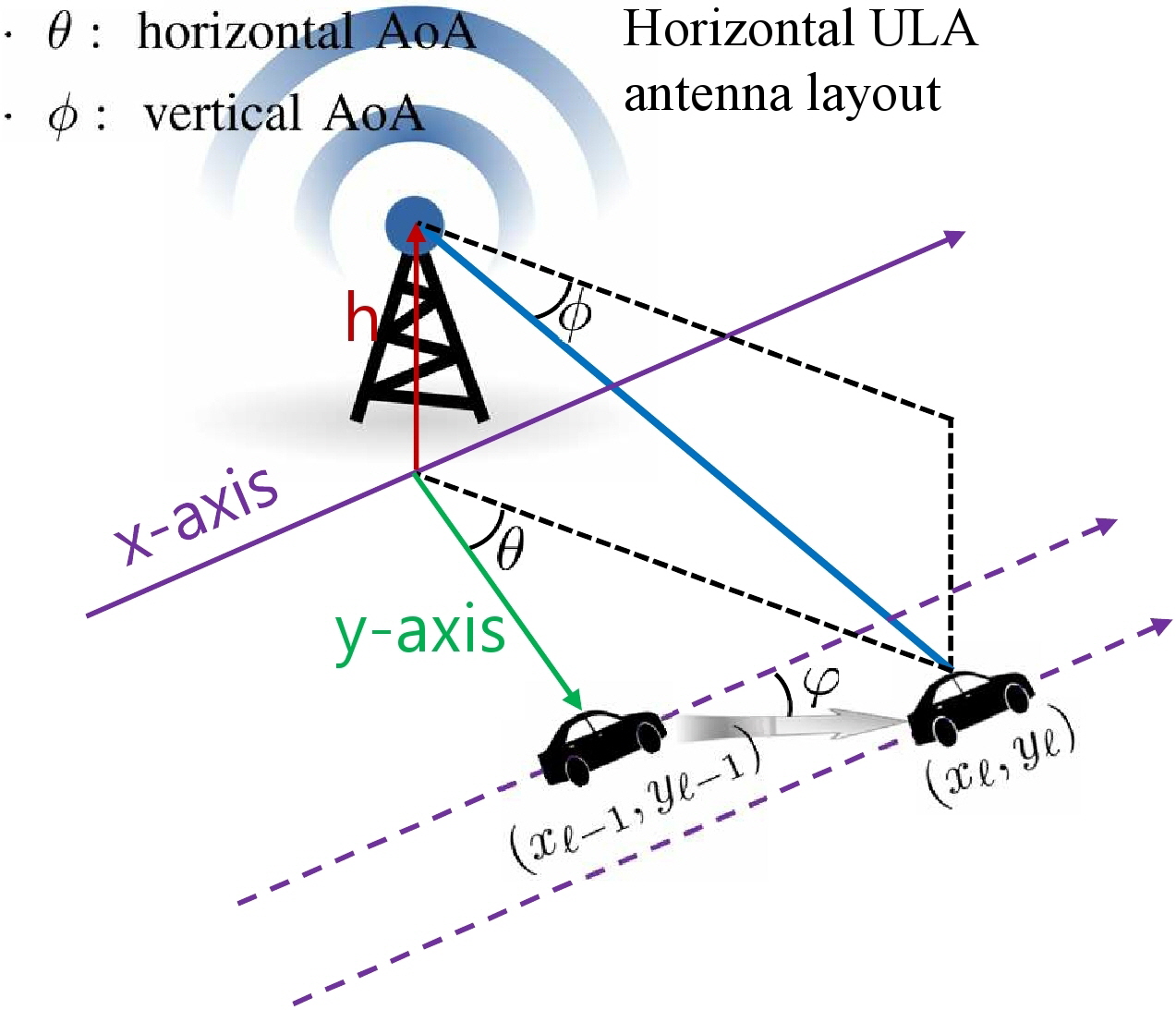}}
\caption{V2I communication system overview.}
\label{fig:system_overview}
\end{figure}

With a single-beam channel model at mmWave bands, the channel vector  can be represented as a function of the beam direction of the dominant radio path $\psi_{\ell}$, such that $\bh_{\ell}= \bh(\psi_{\ell})$.
At discrete time $\ell$, the input-output expression for downlink transmissions is then rewritten as
\begin{align*}
r_{\ell}^{\mathrm{down}}=\sqrt{\rho_{\ell}}  (\beta_{\ell}  \bd_M(\psi_{\ell}))^H \bc_{\ell} s +  n_{\ell}^{\mathrm{down}},
\end{align*}
where $\beta_{\ell} \sim \mathcal{CN}(0,1)$ denotes the small-scale channel fading parameter.
If we consider a uniform linear array (ULA) layout, then the array response vector can be modeled as $\bd_{M}(\psi_{\ell})=[1, e^{j \psi_{\ell}}, \cdots, e^{j (M-1) \psi_{\ell}}]^T \in \mathbb{C}^M$.

The transmitter {mounted on RSU} sends the data symbol to a vehicle {in} the traffic lane.
As depicted in Fig. \ref{fig:system_overview}, the height of RSU is given by $h$, distance from RSU to a vehicle (on the $y$-axis) is defined by $y_{\ell}$, and vehicle position at discrete time $\ell$ in the moving direction ($x$-axis), is denoted as $x_{\ell}$.
Based on the road layout, the horizontal angle of departure (AoD) is defined as $\sin \theta= {x}({x^2+y^2})^{-\frac{1}{2}}$, and the vertical AoD is defined as
$\cos \phi= (x^2+y^2)^{\frac{1}{2}}(x^2+y^2+h^2)^{-\frac{1}{2}}$.
For the ULA scenario, the spatial frequency in the angular domain is then written as
\begin{align}
\label{eq:spatial-geographic}
\psi= \pi \sin\theta \cos  \phi = \pi x(x^2+y^2+h^2)^{-\frac{1}{2}} \doteq \mathrm{T}(x,y).
\end{align}

In this paper, we make the following assumptions:
\begin{itemize}
\item Initial position and velocity information of the vehicle {$(x_0,y_0,v_0)$ is} reported to RSU via a feedback link.
\item Large-scale and small-scale channel fading parameters $(\rho_{\ell},\beta_{\ell})$ {are} known at RSU.
\end{itemize}

\section{{Vehicle Tracking} Algorithm}
\label{sec:beam_tracking_algorithm}

The direction of the dominant radio path from RSU to the vehicles changes significantly due to their high-mobility.
Thus, it is necessary to quickly find the proper beamformer to support high-mobility scenarios in vehicular communication networks.
The channel direction to the vehicle can be obtained by providing RSU with immediate feedback on the vehicle motion states, including position and velocity.
However, the feedback-assisted state tracking approach may lead to a significant increase in the control signaling overhead in wireless networks.
Considering a high-mobility scenario, it is not practical to frequently communicate vehicle motion states via feedback links.
We develop a {vehicle tracking} algorithm using an uplink channel sounding to estimate vehicle motion states without relying on a direct feedback process.

\subsubsection{Linear state prediction}
The state vector of a moving vehicle is defined by, $\bt_{\ell}=[x_{\ell}, y_{\ell}, v_{\ell}]^T \in \mathbb{R}^3$, where $x_{\ell}$, $y_{\ell}$, and  $v_{\ell}$, denote the positions on the $x$-axis, $y$-axis, and velocity of the vehicle at time $\ell$, respectively.
The linear state prediction model at discrete time $\ell$ with step $T_s$ is modeled by
\begin{align}
\label{eq:predict}
\bt_{\ell}= \bA \bt_{\ell-1} + \bb \alpha + \bc_{\ell-1} ,
\end{align}
where $\bA=\Bigg[
             \begin{array}{ccc}
               1 & 0 &T_s \cos\varphi \\
               0 & 1 & T_s   \sin\varphi\\
               0 & 0 & 1   \\
             \end{array}
           \Bigg]
 \in \mathbb{R}^{3 \times 3}$ is the state transition matrix,  $\bb=\big[\frac{T_s^2}{2}\cos\varphi, \frac{T_s^2}{2}\sin\varphi, T_s\big]^T \in \mathbb{R}^{3}$ is the acceleration transition vector with the covariance matrix $\bQ_{\alpha}=\bb\bb^T \sigma_{\alpha}^2$, as in  \cite{Ref_Ken18}.
Similar to \cite{Ref_Sha19,Ref_Lar19}, the error transition vector is modeled using $\bc_{\ell-1}  \sim \mathcal{N}(\mathbf{0}_3, \bQ_{\omega})$   with the covariance matrix $\bQ_{\omega}=\mathrm{diag}[T_s^2\sigma_{\omega}^2\cos^2\varphi, T_s^2\sigma_{\omega}^2\sin^2\varphi, \sigma_{\omega}^2]$, where $\sigma_{\omega}^2$ is the variance of the error parameter.
Note that $\varphi \in [-\pi/2,\pi/2]$ denotes the vehicle steering angle and $\alpha \sim \mathcal{N}(0,\sigma_{\alpha}^2)$ denotes the acceleration parameter that follows Gaussian random process.
In each coherence period, the acceleration parameter is assumed to be stationary, whereas the error parameter is assumed to be dynamic.
Furthermore, we assume that the acceleration parameter  is independent of the error parameter.

\subsubsection{State tracking based on uplink sounding}
RSU only knows the state transition matrix  and the initial state vector\footnote{An initial state vector, ${\bt}_{0}=[x_0, y_0, v_0]^T$, is fed back to RSU at the beginning of a vehicle tracking process. An initial feedback error is modelled by ${\bee}_{\epsilon}=\epsilon{\bt}_{0}$, where $\epsilon \sim N(0,\sigma_{\epsilon}^2)$ denotes the feedback error parameter.}, so the uncertainty of the predicted state vector will gradually increase, without correcting the error transition vector in (\ref{eq:predict}).
It is necessary to predict the acceleration parameter and the error parameter in order to update the state vector precisely.
We focus on estimating two random parameters that affect the  motion state vector using uplink channel sounding signals.

Assuming a time division duplexing architecture, the uplink channel sounding process  for discrete time $\ell$ is
\begin{align}
\label{eq:input_output_uplink}
r_{\ell}=\bz_{\ell}\bh(\psi_{\ell}) + n_{\ell},
\end{align}
where  $\bz_{\ell}  \in \mathbb{C}^{1 \times M}$ is the unit-norm combiner at RSU, $r_{\ell} \in \mathbb{C}$ is the combined baseband signal, $\bh(\psi_{\ell})=\beta_{\ell}\bd_M(\psi_{\ell}) \in \mathbb{C}^{M}$ is the uplink channel vector, and $n_{\ell}=\bz_{\ell}\bn_{\ell} \sim \mathcal{CN}(0,\rho_{\ell}^{-1})$ is the combined noise signal that hinders the channel sounding.
The variables in (\ref{eq:input_output_uplink}) can be reformulated using real and imaginary parts, such that $r_{\ell}=r_{\ell}^{\textrm{re}}+i r_{\ell}^{\textrm{im}}$, $\bz_{\ell}=\bz_{\ell}^{\textrm{re}}+i \bz_{\ell}^{\textrm{im}}$,  $\beta_{\ell}=\beta_{\ell}^{\textrm{re}}+i \beta_{\ell}^{\textrm{im}}$, and $n_{\ell}=n_{\ell}^{\textrm{re}}+i n_{\ell}^{\textrm{im}}$.
Furthermore, the array response vector is rewritten as $\bd_M(\psi_{\ell})=\bd_{M}^{\textrm{re}}(\psi_{\ell})+i \bd_{M}^{\textrm{im}}(\psi_{\ell})$, where the real and imaginary parts of the array vector are
\begin{align}
\label{eq:array_re}
\bd_{M}^{\textrm{re}}(\psi_{\ell})&=\big[\cos(0),\cos(\psi_{\ell}),\cdots, \cos((M-1)\psi_{\ell})\big]^T,
\\
\label{eq:array_im}
\bd_{M}^{\textrm{im}}(\psi_{\ell})&=\big[\sin(0),\sin(\psi_{\ell}),\cdots, \sin((M-1)\psi_{\ell})\big]^T.
\end{align}

A Kalman filter estimates the state vector by {iteratively} applying two steps: {a} dynamic update based on the state prediction model in (\ref{eq:predict}); and {a} measurement update  using the uplink sounding sample in (\ref{eq:input_output_uplink}).
The state prediction process updating position and velocity information is defined in the real domain, whereas the uplink sounding process  in (\ref{eq:input_output_uplink}) is defined in the complex domain.
During the measurement update process, an innovation term that contains new information about the state vector is added to the previous state vector.
However, since measurement noise {hinders} accurate estimation, the complex noise, which makes the state vector complex, would remain in the innovation term.

To conduct the state prediction and update processes in the same domain, the uplink channel sounding process in (\ref{eq:input_output_uplink}) is rewritten in the real domain, such that
\begin{align}
\label{eq:sounding_real}
\tilde{\br}_{\ell}= \tilde{\bZ}_{\ell} \tilde{\bh}(\psi_{\ell}) + \tilde{\bn}_{\ell},
\end{align}
where all variables in {the} real domain are reformulated as {follows:} $\tilde{\br}_{\ell}=[r_{\ell}^{\textrm{re}},~ r_{\ell}^{\textrm{im}}]^T \in \mathbb{R}^2$, $\tilde{\bZ}_{\ell} = \Big[
             \begin{array}{cc}
               \bz_{\ell}^{\textrm{re}} & -\bz_{\ell}^{\textrm{im}} \\
               \bz_{\ell}^{\textrm{im}} & \bz_{\ell}^{\textrm{re}} \\
             \end{array}
           \Big] \in \mathbb{R}^{2 \times 2M}$, and $\tilde{\bn}_{\ell} =[n_{\ell}^{\textrm{re}},~ n_{\ell}^{\textrm{im}}]^T \in \mathbb{R}^{2}$.
The channel vector is reformulated as
\begin{align*}
\tilde{\bh}(\psi_{\ell})=\bigg[
             \begin{array}{cc}
                \bh^{\textrm{re}}(\psi_{\ell})  \\
              \bh^{\textrm{im}}(\psi_{\ell})  \\
             \end{array}
           \bigg]=
\bigg[
             \begin{array}{cc}
               \beta_{\ell}^{\textrm{re}} \bd_{M}^{\textrm{re}}(\psi_{\ell}) - \beta_{\ell}^{\textrm{im}}\bd_{M}^{\textrm{im}}(\psi_{\ell})  \\
              \beta_{\ell}^{\textrm{im}} \bd_{M}^{\textrm{re}}(\psi_{\ell}) + \beta_{\ell}^{\textrm{re}}\bd_{M}^{\textrm{im}}(\psi_{\ell})  \\
             \end{array}
           \bigg].
\end{align*}

The uplink  channel  sounding in (\ref{eq:sounding_real}) is represented by {a} simplified  single-input multiple-output channel vector.
The simplified channel vector is defined using the array response vectors in (\ref{eq:array_re}) and (\ref{eq:array_im}),\ which are the nonlinear functions of the beam direction (spatial frequency).
Furthermore, the beam direction is defined based on  equation   (\ref{eq:spatial-geographic}), which is given as a nonlinear function of the position variables.
Thus, the uplink channel sounding is given as a nonlinear function of the position variables.

The Kalman filter that is typically discussed directly applies only to linear systems.
Nonlinearity exists between the uplink sounding and vehicle position in our case.
Therefore, the uplink channel sounding expression needs to be linearized.
As depicted in Fig. \ref{fig:system_overview}, the beam direction (spatial frequency) at time $\ell$ is defined as a function of the position variable in the state vector, such that $\psi_{\ell}=  \pi x_{\ell}(x_{\ell}^2+y_{\ell}^2+h^2)^{-\frac{1}{2}} \doteq \mathrm{g}(\bt_{\ell})$, and the channel vector is then redefined as $\tilde{\bh}(\mathrm{g}({\bt}_{\ell}))$.
At time $\ell$, the first-order Taylor series approximation of the channel vector {for} {the} predicted state vector $\hat{\bt}_{\ell | \ell-1}$   is given by
\begin{align}
\label{eq:approx_channel}
\tilde{\bh}\big(\mathrm{g}({\bt}_{\ell})\big) \simeq \tilde{\bh}\big(\mathrm{g}(\hat{\bt}_{\ell | \ell-1})\big) + \tilde{\bD}_{\ell|\ell-1} ({\bt}_{\ell }-\hat{\bt}_{\ell | \ell-1}),
\end{align}
where the Jacobian matrix of the channel vector,  $\tilde{\bD}_{\ell|\ell-1} = \frac{\partial \bh(\mathrm{g}(\bt))}{\partial \bt}\big|_{\bt=\hat{\bt}_{\ell|\ell-1}}$, is derived from  {Proposition \ref{prop:01}.}

\begin{prop}
\label{prop:01}
{The} Jacobian matrix of the channel vector is
\begin{align*}
\tilde{\bD}_{\ell|\ell-1} &= \bigg[ \begin{array}{cc}
             \dot{\bh}^{\textrm{re}}(\mathrm{g}(\hat{\bt}_{\ell|\ell-1}))      \\
             \dot{\bh}^{\textrm{im}}(\mathrm{g}(\hat{\bt}_{\ell|\ell-1}))   \\
             \end{array}\bigg]\dot{ \mathrm{g} } (\hat{\bt}_{\ell|\ell-1}),
\end{align*}
where $\dot{ \mathrm{g} } (\hat{\bt}_{\ell|\ell-1}) \doteq \frac{\pi[(\hat{y}_{\ell|\ell-1}^2+h^2), -\hat{x}_{\ell|\ell-1}\hat{y}_{\ell|\ell-1} , \cos\varphi(\hat{y}_{\ell|\ell-1}^2+h^2)T_s]}{(\hat{x}_{\ell|\ell-1}^2+\hat{y}_{\ell|\ell-1}^2+h^2)^{\frac{3}{2}}}$, $\dot{\bh}^{\textrm{re}}(\psi_{\ell}) \doteq \frac{\partial  \bh^{\textrm{re}}(\psi)}{\partial \psi}\Big|_{\psi=\psi_{\ell}} = \beta_{\ell}^{\textrm{re}} \dot{\bd}_{M}^{\textrm{re}}(\psi_{\ell}) - \beta_{\ell}^{\textrm{im}}\dot{\bd}_{M}^{\textrm{im}}(\psi_{\ell})$, and $\dot{\bh}^{\textrm{im}}(\psi_{\ell}) \doteq \frac{\partial  \bh^{\textrm{im}}(\psi)}{\partial \psi} \Big|_{\psi=\psi_{\ell}}= \beta_{\ell}^{\textrm{im}} \dot{\bd}_{M}^{\textrm{re}}(\psi_{\ell}) + \beta_{\ell}^{\textrm{re}}\dot{\bd}_{M}^{\textrm{im}}(\psi_{\ell})$.
\end{prop}
\begin{IEEEproof}
The Jacobian matrix of the simplified channel vector is obtained by solving the equation $\tilde{\bD}_{\ell|\ell-1} = \frac{\partial \bh(\mathrm{g}(\bt))}{\partial \bt}\big|_{\bt=\hat{\bt}_{\ell|\ell-1}}$.
Based on the chain rule, the partial derivative of the channel vector {in} the beam direction, $\hat{\bt}_{\ell|\ell-1}$, is divided into two parts, such that $\tilde{\bD}_{\ell|\ell-1}  =\frac{\partial \tilde{\bh}(\psi)}{\partial \psi}\big|_{\psi = \mathrm{g}(\hat{\bt}_{\ell|\ell-1})} \frac{\partial  \mathrm{g}(\bt) }{\partial \bt}\big|_{\bt=\hat{\bt}_{\ell|\ell-1}}$.
The first part is derived as
\begin{align*}
\frac{\partial \tilde{\bh}(\psi)}{\partial \psi}\bigg|_{\psi=\psi_{\ell}} =\Bigg[
             \begin{array}{cc}
               \beta_{\ell}^{\textrm{re}} \dot{\bd}_{M}^{\textrm{re}}(\psi_{\ell}) - \beta_{\ell}^{\textrm{im}}\dot{\bd}_{M}^{\textrm{im}}(\psi_{\ell})  \\
               \beta_{\ell}^{\textrm{im}} \dot{\bd}_{M}^{\textrm{re}}(\psi_{\ell}) + \beta_{\ell}^{\textrm{re}}\dot{\bd}_{M}^{\textrm{im}}(\psi_{\ell})  \\
             \end{array}
           \Bigg],
\end{align*}
where the partial {derivatives} of the real and imaginary parts of the array response vector {are}
\begin{align*}
\dot{\bd}_{M}^{\textrm{re}}(\psi_{\ell})&=\big[0,-\sin(\psi_{\ell}),\cdots, -(M-1)\sin((M-1)\psi_{\ell})\big]^T,
\\
\dot{\bd}_{M}^{\textrm{im}}(\psi_{\ell})&=\big[0,\cos(\psi_{\ell}),\cdots, (M-1)\cos((M-1)\psi_{\ell})\big]^T.
\end{align*}
Furthermore, the second part is derived as
\begin{align*}
\frac{\partial  \mathrm{g}(\bt) }{\partial \bt}&=\bigg[ \frac{\partial g(\bt) }{\partial x},\frac{\partial  g(\bt) }{\partial y} ,\frac{\partial g(\bt) }{\partial v}   \bigg] \bigg|_{\bt=[x,y,v]^T}
\\
& \stackrel{(a)} \simeq  \frac{\pi[(y^2+h^2), -xy ,(y^2+h^2)T_s  \cos\varphi]}{(x^2+y^2+h^2)^{\frac{3}{2}}} ,
\end{align*}
where $(a)$ is derived because $\frac{\partial x}{\partial v} \simeq  T_s \cos \varphi$.
\end{IEEEproof}

By substituting the approximated channel vector in (\ref{eq:approx_channel}) into (\ref{eq:sounding_real}), the input-output expression for the uplink channel sounding can be approximated by
\begin{align}
\label{eq:linearized_io}
\tilde{\br}_{\ell} \simeq \tilde{\bZ}_{\ell} \tilde{\bD}_{\ell|\ell-1}{\bt}_{\ell} + \tilde{\bZ}_{\ell}  \tilde{\bh}\big(\mathrm{g}(\hat{\bt}_{\ell|\ell-1})\big) - \tilde{\bZ}_{\ell} \tilde{\bD}_{\ell|\ell-1}\hat{\bt}_{\ell|\ell-1}   + \tilde{\bn}_{\ell}.
\end{align}
Similar to \cite{Ref_Kay93,Ref_Ter08,Ref_Sha19}, we can develop EKF by putting the rewritten equation in (\ref{eq:linearized_io}) into the Kalman filter, instead of exploiting the nonlinear channel sounding equation in (\ref{eq:sounding_real}).

Based on the state transition model in (\ref{eq:predict}), the state vector and covariance matrix can be updated as $\hat{\bt}_{\ell|\ell-1}=\bA\hat{\bt}_{\ell-1}$ and $\bQ_{\ell|\ell-1}=\bA \bQ_{\ell-1} \bA^T + \bQ_{\alpha}+\bQ_{\omega}$, respectively.
Since the acceleration and error parameters cannot be corrected in the state prediction process, the uncertainty of the state vector gradually increases.
To correct the prediction errors, the state vector and the covariance matrix should be updated based on the EKF algorithm.
The updated state vector is given by
\begin{align*}
\hat{\bt}_{\ell}={\hat{\bt}_{\ell|\ell-1}} +\tilde{\bK}_{\ell}\big(\tilde{\br}_{\ell}-\tilde{\bZ}_{\ell}  \tilde{\bh}(\mathrm{g}(\hat{\bt}_{\ell|\ell-1}))\big),
\end{align*}
and the updated covariance matrix is given by
\begin{align*}
\bQ_{\ell}=\big(\bI_3- \tilde{\bK}_{\ell} \tilde{\bZ}_{\ell} \tilde{\bD}_{\ell|\ell-1}\big)\bQ_{\ell|\ell-1},
\end{align*}
{in which} the Kalman gain matrix is defined by
\begin{align}
\label{eq:kalman}
\tilde{\bK}_{\ell}=\bQ_{\ell|\ell-1} \tilde{\bD}_{\ell|\ell-1}^T \tilde{\bZ}_{\ell}^T  \Big(\tilde{\bZ}_{\ell} \tilde{\bD}_{\ell|\ell-1} \bQ_{\ell|\ell-1} \tilde{\bD}_{\ell|\ell-1}^T\tilde{\bZ}_{\ell}^T +\frac{\bI_2}{2\rho_{\ell}}  \Big)^{-1}.
\end{align}
The Kalman gain matrix is designed to optimize the state prediction and update process by considering both the  state transition process  and approximated channel sounding expression together \cite{Ref_Kay93,Ref_Ter08}.

\subsubsection{Combining vector design for uplink channel sounding}
We develop a method for computing the optimal combiner for the uplink sounding process.
The objective of the  combining process is to design a combiner that maximizes the overall state estimation performance of the EKF algorithm.
As studied in \cite{Ref_Cho14,Ref_Lar19}, we aim to compute an optimal  combiner at RSU that minimizes the trace of the covariance matrix.
The combiner  is constructed by computing the real and imaginary parts of the combined vector in the real domain, such that
\begin{align}
\nonumber
&(\bz_{\ell}^{\textrm{re}},\bz_{\ell}^{\textrm{im}}) = \argmin_{\bar{\bz}_{\ell}^{\textrm{re}},\bar{\bz}_{\ell}^{\textrm{im}}} \mathrm{Tr}\{\bQ_{\ell}\}
\\
\nonumber
&= \argmax_{\bar{\bz}_{\ell}^{\textrm{re}},\bar{\bz}_{\ell}^{\textrm{im}} } \mathrm{Tr}\{ \tilde{\bK}_{\ell} \bar{\bZ}_{\ell} \tilde{\bD}_{\ell|\ell-1}\bQ_{\ell|\ell-1}\}
\\
\nonumber
&= \argmax_{\bar{\bz}_{\ell}^{\textrm{re}},\bar{\bz}_{\ell}^{\textrm{im}} \in \mathbb{R}^{1 \times M}} \mathrm{Tr}\{\bQ_{\ell|\ell-1}  \tilde{\bD}_{\ell|\ell-1}^T\bar{\bZ}_{\ell}^T \mathbf{\Lambda}_{\ell}^{-1} \bar{\bZ}_{\ell} \tilde{\bD}_{\ell|\ell-1}\bQ_{\ell|\ell-1}\}
\\
\label{eq:opt_pro_real}
&  \stackrel{(a)} = \argmax_{\bar{\bz}_{\ell}^{\textrm{re}},\bar{\bz}_{\ell}^{\textrm{im}}\in \mathbb{R}^{1 \times M} } \mathrm{Tr}\{ \mathbf{\Lambda}_{\ell}^{-1} \bar{\bZ}_{\ell} \tilde{\bD}_{\ell|\ell-1}\bQ_{\ell|\ell-1} ^2\tilde{\bD}_{\ell|\ell-1}^T \bar{\bZ}_{\ell}^T \},
\end{align}
where $\mathbf{\Lambda}_{\ell} \doteq (\bar{\bZ}_{\ell} \tilde{\bD}_{\ell|\ell-1} \bQ_{\ell|\ell-1}  \tilde{\bD}_{\ell|\ell-1}^T\bar{\bZ}_{\ell}^T + {\bI}_2/{(2\rho_{\ell})})$, the combining matrix in the real domain is defined by $\bar{\bZ}_{\ell} = \Big[
             \begin{array}{cc}
               \bar{\bz}_{\ell}^{\textrm{re}} & -\bar{\bz}_{\ell}^{\textrm{im}} \\
               \bar{\bz}_{\ell}^{\textrm{im}} & \bar{\bz}_{\ell}^{\textrm{re}} \\
             \end{array}
           \Big]$, and $(a)$ is derived {as} $\mathrm{Tr}\{\bA\bB\}=\mathrm{Tr}\{\bB\bA\}$ \cite{Ref_Pet12}.

However, in the real domain, it is not easy to jointly compute the real and imaginary parts of the combining vector.
Therefore, we rewrite the  problem in (\ref{eq:opt_pro_real}) as an optimization problem in the complex domain, such that
\begin{align}
\nonumber
\bz_{\ell} &= \argmax_{\bar{\bz}\in \mathbb{C}^{1 \times M}} \frac{\bar{\bz}\big(\bD_{\ell|\ell-1}\bQ_{\ell|\ell-1}^2\bD_{\ell|\ell-1}^H\big)\bar{\bz}^H}{\bar{\bz}\big(\bD_{\ell|\ell-1}\bQ_{\ell|\ell-1}\bD_{\ell|\ell-1}^H+{\bI_M}/{\rho_{\ell}}\big)\bar{\bz}^H}
\\
\nonumber
& \stackrel{(a)} = \mathfrak{v}\big\{ \big(\bD_{\ell|\ell-1} \bQ_{\ell|\ell-1} \bD_{\ell|\ell-1}^{H} + {\bI_M}/{\rho_{\ell}} \big)^{-1}
\\
\label{eq:opt_com}
&~~~~~~~~~~~~~~~~~~~~~~~~~~~~~~~~\big(\bD_{\ell|\ell-1} \bQ_{\ell|\ell-1}^2 \bD_{\ell|\ell-1}^H\big)\big\},
\end{align}
where $\bD_{\ell|\ell-1} = \big[ \frac{\partial \bh^{\textrm{re}}(\psi)}{\partial \psi}+i  \frac{\partial \bh^{\textrm{im}}(\psi)}{\partial \psi}
\big]\big|_{\psi=\mathrm{g}(\hat{\bt}_{\ell|\ell-1})}\dot{ \mathrm{g} } (\bt)\big|_{\bt=\hat{\bt}_{\ell|\ell-1}}$,  $\bar{\bz}=\bar{\bz}_{\ell}^{\textrm{re}}+i\bar{\bz}_{\ell}^{\textrm{im}}$ and $(a)$ {are} derived based on {the} Rayleigh quotient algorithm \cite{Ref_Bor98}.
The hybrid beamforming system at RSU might not be able to construct the optimal combiner in (\ref{eq:opt_com}), due to an equal gain constraint on the analog beamforming hardware \cite{Ref_Aya14,Ref_Alk14}.
A complex combiner should be  reconstructed to satisfy the power constraint of the hybrid beamforming setup.
To generate a complex combiner close to the optimal combiner, a set of analog beam steering vectors and a digital weight vector are constructed based on the orthogonal matching pursuit algorithm.
Please refer to \cite{Ref_Reb02,Ref_Aya14,Ref_Alk14,Ref_Son17} for the hybrid combiner design problem.
Finally, the complex combiner ${\bz}_{\ell}={\bz}_{\ell}^{\textrm{re}}+i{\bz}_{\ell}^{\textrm{im}}$  is then  reformulated in the {real domain as} $\tilde{\bZ}_{\ell} = \Big[
             \begin{array}{cc}
               \bz_{\ell}^{\textrm{re}} & -\bz_{\ell}^{\textrm{im}} \\
               \bz_{\ell}^{\textrm{im}} & \bz_{\ell}^{\textrm{re}} \\
             \end{array}
           \Big]$, and the matrix {is} used to rewrite the uplink sounding signal, as in (\ref{eq:sounding_real}).

\subsubsection{Acceleration parameter estimation}
The proposed vehicle tracking algorithm aims to correct the errors between the actual state vector and predicted state vector caused by the acceleration and error variables.
The vehicle tracking algorithm sequentially corrects the errors with the aid of the uplink channel sounding process.
If we estimate the acceleration parameter successfully, then we can focus on correcting the  errors via  Kalman filtering, because the acceleration parameter is assumed to be stationary in each coherence time.
Therefore, we develop a method for estimating the acceleration parameter using the updated state vectors.

We define the long-term state transition equation, which represents the relationship between {the} initial state vector  ${\bt}_{0}$ and current state vector ${{\bt}_{\ell}}$ at time $\ell$, such that
\begin{align}
\nonumber
{{\bt}_{\ell}}   &=  \bA^{\ell}  {\bt}_{0}+ \sum_{\tau=1}^{\ell} \bA^{\tau-1}\bb \alpha +\sum_{\tau=1}^{\ell} \bA^{\tau-1} {\bc_{\tau-1}}
\\
\label{eq:acc_par}
&= \bA^{\ell}  {\bt}_{0} + \tilde{\bb}_{\ell} \alpha + \tilde{\bc}_{\ell},
\end{align}
where the modified acceleration transition vector is given by $\tilde{\bb}_{\ell}\doteq \sum_{\tau=1}^{\ell} \bA^{\tau-1}\bb=[(\ell T_s)^2/2 \cos\varphi, (\ell T_s)^2/2 \sin\varphi, \ell T_s]^T$, and the  modified error vector is $\tilde{\bc}_{\ell} \doteq \sum_{\tau=1}^{\ell} \bA^{\tau-1} {\bc_{\tau-1}}$.

The state vector  can be modeled using the estimated state vector and its corresponding covariance matrix, such as
\begin{align}
\label{eq:mod_vec}
{{\bt}_{\ell}}   = \hat{\bt}_{\ell}   +  \bee_{\ell},
\end{align}
where the error term is modeled based on {the} \textit{Correlated Grassmannian} algorithm  such as $\bee_{\ell} \doteq {\bQ_{\ell}^{\frac{1}{2}} \bw}/{\|\bQ_{\ell}^{\frac{1}{2}} \bw\|_2}$ with {a} random vector $\bw \sim \mathcal{N}(\mathbf{0}_3,\bI_3)$.
By {substituting} the  vector in (\ref{eq:mod_vec}) into  state equation (\ref{eq:acc_par}), we  rewrite the state model in (\ref{eq:acc_par}) for the acceleration parameter estimation,  such that
\begin{align*}
\nonumber
\tilde{\bt}_{\ell} &\doteq  \hat{\bt}_{\ell}   - \bA^{\ell}  {\bt}_{0}= \tilde{\bb}_{\ell} \alpha + \tilde{\bc}_{\ell} - \bee_{\ell},
\end{align*}
where the error vector can be modeled using {a} normal distribution, such that $\tilde{\bc}_{\ell} \sim \mathcal{N}\big(\mathbf{0}_3 , \bC_{\ell} \big)$.
The covariance matrix is derived as
\begin{align*}
\bC_{\ell} &\doteq \mathrm{E}\big[ \tilde{\bc}_{\ell} \tilde{\bc}_{\ell}^T \big]
=\sum_{\tau=1}^{\ell} \bA^{\tau-1} \bQ_{\omega}(\bA^{\tau-1})^T    \in \mathbb{R}^{3 \times 3}.
\end{align*}
Based on the  minimum variance unbiased (MVU) estimator in \cite{Ref_Kay93}, the acceleration parameter is estimated as
\begin{align}
\label{eq:acc_estimation}
\hat{\alpha}&=\frac{\tilde{\bb}_{\ell}^T (\bC_{\ell}+\bQ_{\ell})^{-1} \tilde{\bt}_{\ell}}{\tilde{\bb}_{\ell}^T (\bC_{\ell}+\bQ_{\ell})^{-1} \tilde{\bb}_{\ell}}=\alpha+\frac{\tilde{\bb}_{\ell}^T (\bC_{\ell}+\bQ_{\ell})^{-1} (\tilde{\bc}_{\ell}-{\bee}_{\ell})}{\tilde{\bb}_{\ell}^\mathrm{T}(\bC_{\ell}+\bQ_{\ell})^{-1} \tilde{\bb}_{\ell}}.
\end{align}

\begin{algorithm}
  \caption{State tracking algorithm using the EKF algorithm}
  \label{Al:01}
  \begin{algorithmic}
\State \textbf{{Initialization}}
\State ~1:~~{Initial state vector},~~{${\bt}_{0}=[x_0, y_0, v_0]^T,~~\hat{\bt}_{0}={\bt}_{0}+{\bee}_{\epsilon}$}
\State ~2:~~{Initial covariance matrix},~~{$\bQ_{0}=\sigma_{\epsilon}^2({\bt}_{0}{\bt}_{0}^T)$} 
\State \textbf{{Linear state prediction}}
\State ~3:~~{Predict state vector}
\State ~~~~~$\hat{\bt}_{\ell|\ell-1}= \bA \hat{\bt}_{\ell-1}$,~~w/o estimated $\hat{\alpha}$
\State ~~~~~$\hat{\bt}_{\ell|\ell-1}= \bA \hat{\bt}_{\ell-1}+\bb\hat{\alpha}$,~~w/ estimated $\hat{\alpha}$
\State ~4:~~{Update covariance matrix}
\State ~~~~~$\bQ_{\ell|\ell-1}=\bA \bQ_{\ell-1} \bA^T + \bQ_{\alpha}+\bQ_{\omega}$,~~w/o estimated $\hat{\alpha}$
\State ~~~~~$\bQ_{\ell|\ell-1}=\bA \bQ_{\ell-1} \bA^T + \bQ_{\omega}$,~~w/ estimated $\hat{\alpha}$
\State \textbf{{Uplink channel sounding}}
\State ~5:~~{Compute {combiner at RSU}}
\State ~~~~~$\bz_{\ell}= \mathfrak{v}\big\{ \big(\bD_{\ell|\ell-1} \bQ_{\ell|\ell-1} \bD_{\ell|\ell-1}^{H} + \frac{\bI_M}{\rho_{\ell}} \big)^{-1}$
\State ~~~~~~~~~~~~~~~~~~~~~~~~~~~~~~~~~~~~~~~~~~~~~~~~~~~~~~~~~$\big(\bD_{\ell|\ell-1} \bQ_{\ell|\ell-1}^2 \bD_{\ell|\ell-1}^H\big)\big\}$
\State ~6:~~{Conduct uplink channel sounding}
\State ~~~~~$r_{\ell}=\bz_{\ell}\bh(\psi_{\ell}) + n_{\ell}~\rightarrow~\tilde{\br}_{\ell}= \tilde{\bZ}_{\ell} \tilde{\bh}(\psi_{\ell}) + \tilde{\bn}_{\ell}$
\State \textbf{{State update based on uplink channel sounding}}
\State ~7:~~{Design Kalman gain matrix}
\State ~~~~~$\tilde{\bK}_{\ell}=\bQ_{\ell|\ell-1} \tilde{\bD}_{\ell|\ell-1}^T \tilde{\bZ}_{\ell}^T$
\State ~~~~~~~~~~~~~~~~~~~~~~~~~~~~~~~~~~~~$\big(\tilde{\bZ}_{\ell} \tilde{\bD}_{\ell|\ell-1} \bQ_{\ell|\ell-1} \tilde{\bD}_{\ell|\ell-1}^T\tilde{\bZ}_{\ell}^T +\frac{\bI_2}{2\rho_{\ell}}  \big)^{-1}$
\State ~8:~~{Update state vector}
\State ~~~~~$\hat{\bt}_{\ell}={\hat{\bt}_{\ell|\ell-1}} +\tilde{\bK}_{\ell}\big(\tilde{\br}_{\ell}-\tilde{\bZ}_{\ell}  \tilde{\bh}(\mathrm{g}(\hat{\bt}_{\ell|\ell-1}))\big)$
\State ~9:~~{Update covariance matrix}
\State ~~~~~$\bQ_{\ell}=(\bI_3- \tilde{\bK}_{\ell} \tilde{\bZ}_{\ell} \tilde{\bD}_{\ell|\ell-1})\bQ_{\ell|\ell-1}$
  \end{algorithmic}
\end{algorithm}

We also study the performance of the MVU estimator.
For any unbiased estimator $\check{\alpha}$, the variance is lower-bounded,  such that
\begin{align*}
\mathrm{var}(\check{\alpha})&\geq {\Big(\mathrm{E}\Big[-\frac{\partial^2}{\partial \alpha^2} \mathrm{ln}( p(\tilde{\bt}_{l};\alpha)) \Big]\Big)^{-1}}
\\
&=\Big(\mathrm{E}\Big[\frac{\partial^2}{\partial \alpha^2}\Big[\frac{(\tilde{\bt}_\ell-\tilde{\bb}_\ell\alpha)^{T}(\bC_{\ell}+\bQ_{\ell})^{-1}(\tilde{\bt}_\ell-\tilde{\bb}_\ell\alpha)}{2}
\\
&~~~~~~~~~~~~~~~~~~~~~~~~~~~~~~~~+\mathrm{ln}(2\pi|\bC_{l}+\bQ_{l}|)\Big]\Big]\Big)^{-1}
\\
&=\Big(\mathrm{E}\Big[\frac{\partial}{\partial \alpha}\Big[ {\tilde{\bb}_{\ell}^T (\bC_{\ell}+\bQ_{\ell})^{-1} \tilde{\bt}_{\ell}}
\\
&~~~~~~~~~~~~~~~~~~~~~~~~+\alpha\big({\tilde{\bb}_{\ell}^T (\bC_{\ell}+\bQ_{\ell})^{-1} \tilde{\bb}_{\ell}}\big)  \Big]\Big]\Big)^{-1}
\\
&=\big({\tilde{\bb}_{\ell}^T (\bC_{\ell}+\bQ_{\ell})^{-1} \tilde{\bb}_{\ell}}\big)^{-1}  = 1/\mathrm{I}(\alpha),
\end{align*}
where $\tilde{\bt}_{l} \sim \cN\big(\mathbf{\tilde{b}_{\ell}}\alpha , \bC_{l}+\bQ_{l} \big)$, $\mathrm{I}(\alpha)$ is the Fisher information, and the Cramer-Rao lower bound (CRLB) is given by $1/\mathrm{I}(\alpha)$.

The variance of the MVU estimator in (\ref{eq:acc_estimation}) is calculated as
\begin{align*}
&\mathrm{var}(\hat{\alpha}) = \mathrm{E}\big[(\hat{\alpha}-\mathrm{E}[\hat{\alpha}])(\hat{\alpha}-\mathrm{E}[\hat{\alpha}])^{T}\big]
\\
&\stackrel{(a)} = \mathrm{E}\big[(\hat{\alpha}-\alpha)(\hat{\alpha}-\alpha)^{T}\big]
\\
& = \mathrm{E}\Bigg[\frac{\tilde{\bb}_{\ell}^T (\bC_{\ell}+\bQ_{\ell})^{-1} (\tilde{\bc}_{\ell}-{\bee}_{\ell})(\tilde{\bc}_{\ell}-{\bee}_{\ell})^{T}(\bC_{\ell}+\bQ_{\ell})^{-1}\tilde{\bb}_{\ell}}{\big|\tilde{\bb}_{\ell}^T
(\bC_{\ell}+\bQ_{\ell})^{-1} \tilde{\bb}_{\ell}\big|^{2}}\Bigg]
\\
& \stackrel{(b)} = \frac{\tilde{\bb}_{\ell}^T (\bC_{\ell}+\bQ_{\ell})^{-1}
\mathrm{E}\Big[(\tilde{\bc}_{\ell}-{\bee}_{\ell})(\tilde{\bc}_{\ell}-{\bee}_{\ell})^{T}\Big](\bC_{\ell}+\bQ_{\ell})^{-1}\tilde{\bb}_{\ell}}{\big|\tilde{\bb}_{\ell}^T (\bC_{\ell}+\bQ_{\ell})^{-1} \tilde{\bb}_{\ell}\big|^{2}}
\\
& \stackrel{(c)} =\big({\tilde{\bb}_{\ell}^T (\bC_{\ell}+\bQ_{\ell})^{-1} \tilde{\bb}_{\ell}}\big)^{-1},
\end{align*}
where $(a)$ is derived based on the unbiased estimator, $(b)$ is derived because only $\tilde{\bc}_{\ell}$ and ${\bee}_{\ell}$ are random variables, and $(c)$ is derived because $\mathrm{E}[\tilde{\bc}_{\ell}\tilde{\bc}_{\ell}^{{T}}]=\bC_{\ell}$, $\mathrm{E}[{\bee}_{\ell}{\bee}_{\ell}^{{T}}]=\bQ_{\ell}$, and $\mathrm{E}[\tilde{\bc}_{\ell}{\bee}_{\ell}^{{T}}]=\mathbf{0}_{3 \times 3}$.
The variance of $\check{\alpha}$ is equal to or greater than {the} CRLB, which denotes the variance of {the} proposed estimator, such that $\mathrm{var}(\check{\alpha})\geq 1/\mathrm{I}(\alpha) = \mathrm{var}(\hat{\alpha})$.
We conclude that the MVU estimator in (\ref{eq:acc_estimation}) is efficient because it achieves the CRLB.

\begin{figure}
\centering
\subfigure{\includegraphics[width=0.425\textwidth]{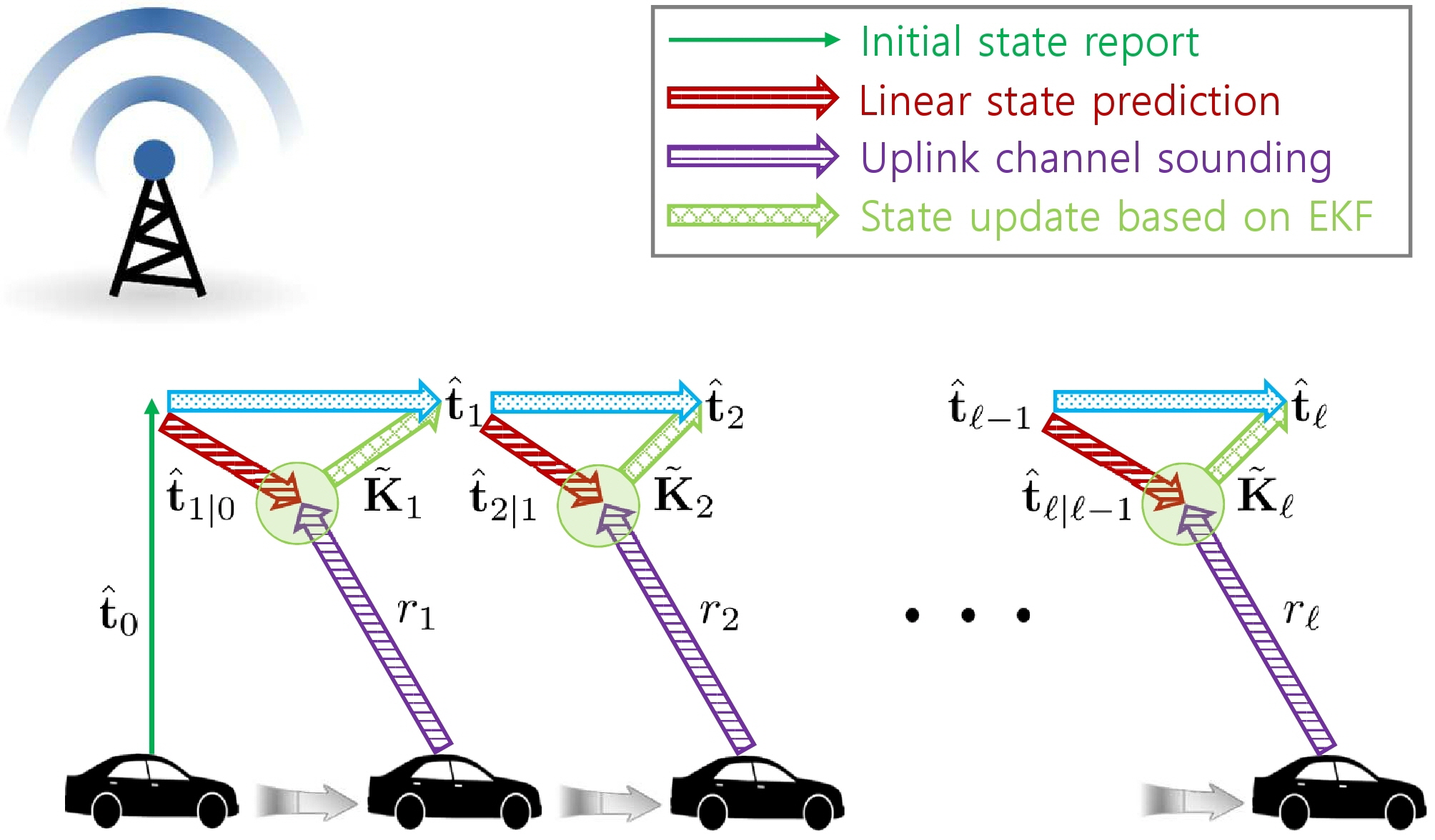}}
\caption{Vehicle tracking algorithm with extended Kalman filter.}
\label{fig:special_01}
\end{figure}

Assuming that the acceleration parameter is properly estimated, we can reduce the randomness/uncertainty in the state prediction process.\footnote{The estimated acceleration parameter, $\hat{\alpha}$, applies to the state prediction process only when the difference between the previous and current estimations is not greater than the predefined threshold parameter, $\alpha_{\mathrm{thres}}$.}
In the proposed EKF algorithm, we focus on correcting the position and velocity changes due to the error parameter, without needing to consider the  changes due to the acceleration of the vehicle,  $\alpha$.
In the state prediction process, the state vector and corresponding covariance matrix can be predicted as
\begin{align*}
\hat{\bt}_{\ell|\ell-1}&= \bA \hat{\bt}_{\ell-1}+\bb\hat{\alpha},
\\
\bQ_{\ell|\ell-1}&=\bA \bQ_{\ell-1} \bA^T + \bQ_{\omega}.
\end{align*}
The transmit combiner for the uplink sounding process can be designed to cover a narrower beam width area, because we only consider the covariance matrix due to the error parameter.
Finally, the EKF algorithm for the state update process is summarized in Algorithm \ref{Al:01} and depicted in Fig. \ref{fig:special_01}.

\begin{figure}
\centering
\subfigure{\includegraphics[width=0.425\textwidth]{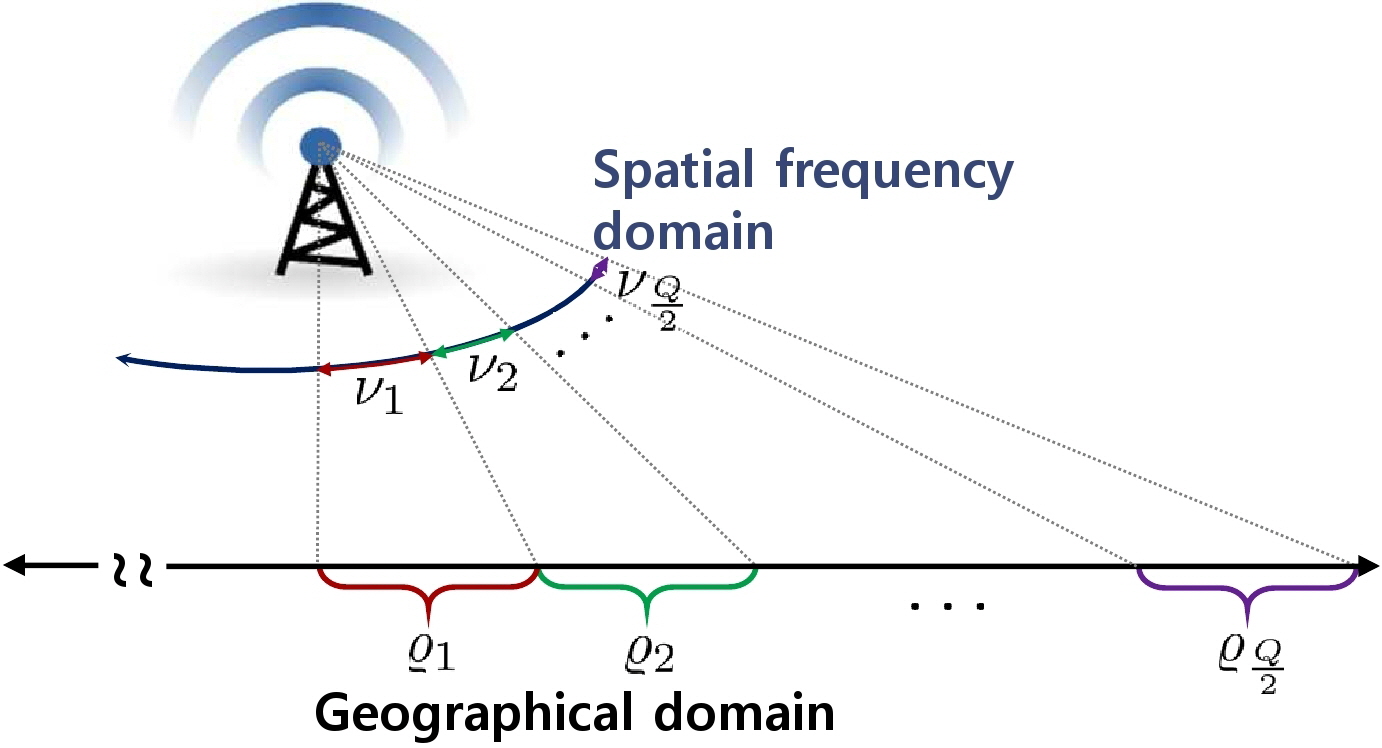}}
\caption{Sub-range transformation between the geographical and spatial frequency domains.}
\label{fig:special_14}
\end{figure}

\section{Downlink Beamformer Design Algorithm}
\label{sec:adaptive_BP}
We design a beamforming codebook by utilizing the knowledge of road structures.
Furthermore, we develop a codeword selection algorithm to obtain a robust beamformer based on the predicted channel information.
Typically, each beamforming codeword  covers a certain beam-region in the spatial frequency domain, which maps onto a geographic area.
The conventional beamformer is designed to generate a beam pattern (BP) within an equally-spaced beam-region  \cite{Ref_Son14_2,Ref_Son17,Ref_Kim20}.
Assuming there are $Q$ codewords $\{\bu_{1},\cdots, \bu_{Q}\}$ in the  codebook, each beam-region is defined to have {the} same length, $|\tilde{\nu}|=\frac{\nu^{\mathrm{ub}}-\nu^{\mathrm{lb}}}{Q}= {\frac{2\nu^{\mathrm{ub}}}{Q}}$, in the spatial frequency domain, where $(\nu^{\mathrm{lb}},\nu^{\mathrm{ub}})$ denotes the left-hand/right-hand edges of the entire beam-region, respectively.

However, the conventional beam-region division approach may not be effective in V2I communications because the distance between RSU and the vehicle on the $y$-axis, $k=\sqrt{y^2+h^2}$, is significantly shorter than the length of the road on the $x$-axis.
As shown in Fig. \ref{fig:special_14},  the difference between the beam width in the spatial frequency domain  and the projected length on the road in the geographical domain varies depending on the vehicle location.
The  beamforming codewords need to be designed by taking the distortion of the projection into account.

\subsubsection{Beam-region division algorithm}
The studies in this sub-section are aimed at   defining the target beam-region of interest for each beamforming codeword.
For a simple analysis, the position on the $y$-axis is assumed to have a fixed value.\footnote{Since we assume that the vehicle does not change lanes, the change of position on the $y$-axis is negligible compared to that on the $x$-axis.}
In the geographical domain, the $q$-th  sub-range is defined by $\varrho_{q}\doteq [ \varrho_{q}^{\mathrm{lb}}, \varrho_{q}^{\mathrm{ub}})$, where $\varrho_{q}^{\mathrm{lb}}$ and $\varrho_{q}^{\mathrm{ub}}$ denote the left-hand edge and the right-hand edge of the sub-range, respectively.
We also denote $\nu_{q} \doteq [\nu_{q}^{\mathrm{lb}}, \nu_{q}^{\mathrm{ub}} )$ as the $q$-th beam-region in the spatial frequency domain, where $\nu_{q}^{\mathrm{lb}}$ and $\nu_{q}^{\mathrm{ub}}$ are the left-hand/right-hand edges of the beam-region, respectively.
The geographic sub-range  can be converted into {the beam-region  $(\nu_{q}\doteq \mathrm{D}[\varrho_{q}])$  and vice versa $(\varrho_{q}\doteq \mathrm{D}^{-1}[\nu_{q}])$.}
For the remainder of this sub-section, we  only consider the positive region, because the opposite region can be  expressed by using symmetry.

Before developing the beam-region division algorithm, we pause to catch up on the conversion process between the geographical and spatial frequency domains.
For further discussion, each geographic sub-range is tentatively defined as having the same length, $|\tilde{\varrho}|=\frac{\varrho^{\mathrm{ub}}-\varrho^{\mathrm{lb}}}{Q}= {\frac{2\varrho^{\mathrm{ub}}}{Q}}$, where $(\varrho^{\mathrm{lb}},\varrho^{\mathrm{ub}})$ {denotes the left-hand/right-hand edges, respectively,} of the geographic range.
This means that all geographic sub-ranges are the {same:} $|\tilde{\varrho}|=|\varrho_{q}|$ for all $q \in \{1,\cdots,Q\}$.
In the spatial frequency domain, however, the transformed beam-region, $\nu_{q} \doteq \mathrm{D}[\varrho_{q}]$ varies due to the nonlinear transformation.

We study the change in the transformed beam-region in the spatial frequency domain.
First, we verify that the beam-region,  $\nu_{q}$, strictly decreases for $q$, because the difference between the derivative of the beam-regions is derived as
\begin{align*}
{\dot{\mathrm{T}}(\varrho_{q}^{\mathrm{ub}},y)-\dot{\mathrm{T}}( \varrho_{q}^{\mathrm{lb}},y)}&=\frac{\pi k^{2}}{(k^{2}+(\varrho_{q}^{\mathrm{ub}})^{2})^\frac{3}{2}}-\frac{\pi k^{2}}{(k^{2}+(\varrho_{q}^{\mathrm{lb}})^{2})^\frac{3}{2}}
\\
& \stackrel{(a)} <0,
\end{align*}
where the nonlinear transformation function, $\mathrm{T}(\cdot)$, is defined in (\ref{eq:spatial-geographic}), $k=\sqrt{y^2+h^2}$, and $(a)$ holds  with $(k^{2}+(\varrho_{q}^{\mathrm{ub}})^{2})^\frac{3}{2}> (k^{2}+(\varrho_{q}^{\mathrm{lb}})^{2})^\frac{3}{2}$ because $(\cdot)^{\frac{3}{2}}$ is {a} strictly increasing function, and $\varrho_{q}^{\mathrm{ub}} > \varrho_{q}^{\mathrm{lb}}$.

Furthermore, we take a closer look at the beam-regions as a function of the position on the road.
The length of the first beam-region in the spatial frequency domain  is greater than that of the equally divided beam-region, such that $|\nu_{1}| > |\tilde{\nu}|$, because
\begin{align*}
|\nu_{1}| - |\tilde{\nu}| & \stackrel{(a)} = { {\mathrm{T}}(\varrho_{1}^{\mathrm{ub}},y) -  {2}{\mathrm{T}}(\varrho^{\mathrm{ub}},y) {Q}^{-1}}
\\
&  \stackrel{(b)} = \frac{\pi|\tilde{\varrho}|}{\big( k^2 + |\tilde{\varrho}|^2 \big)^{\frac{1}{2}}}- \frac{{\pi |\tilde{\varrho}|}}{\big( {k^2 + (|\tilde{\varrho}| Q/2)^2} \big)^{\frac{1}{2}}} \stackrel{(c)} > 0,
\end{align*}
where {$(a)$ holds because $|\tilde{\nu}|= 2{\nu^{\mathrm{ub}}}/{Q}$}, $(b)$ holds  with {$\varrho^{\mathrm{ub}}=|\tilde{\varrho}| Q/2$},  and  $(c)$ holds because $(\cdot)^{\frac{1}{2}}$ is a strictly increasing function {with $Q>2$}.
Furthermore, assuming that the vehicle is far away from the center of the road,  the length of the converted beam-region  is compared to that of the equally divided beam-region in the following Lemma.
\begin{lemma}
\label{lm:01}
Assuming $|x| \gg k$, the length of the equally divided beam-region over the length of the converted beam-region is greater than one, such that ${|\tilde{\nu}|}/{|\nu_{q} |}>1$  when the approximated condition, $|x|>   {\frac{\sqrt{|\varrho |(|\varrho |+kQ)}}{2}}$, holds.
\end{lemma}
\begin{IEEEproof}
The ratio between the length of the equally divided beam-region  and the length of the converted beam-region is approximated by
\begin{align*}
\Delta_{q} &\doteq \frac{|\tilde{\nu}|}{|\nu_{q} |} =\frac{ {2 \pi Q^{-1} }{\big( 1 + \big( \frac{{2k}}{|\tilde{\varrho}| Q} \big)^2   \big)^{-\frac{1}{2}}}}{{\pi}\big({{1 +\big( \frac{k}{\varrho_{q}^{\mathrm{ub}}}\big)^2}}\big)^{-\frac{1}{2}} - {\pi}\big({{1 +\big(\frac{k}{\varrho_{q}^{\mathrm{lb}}}\big)^2}}\big)^{-\frac{1}{2}}}
\\
& \stackrel{(a)} \simeq \frac{{2\pi Q^{-1}} \big({{1 -\frac{k}{{ |\tilde{\varrho}| Q }}}}\big) }{ {\pi}\big({{1 -\frac{k}{2\varrho_{q}^{\mathrm{ub}}}}}\big) - {\pi}\big({{1 -\frac{k}{2\varrho_{q}^{\mathrm{lb}}}}}\big)}={\frac{\varrho_{q}^{\mathrm{ub}} \varrho_{q}^{\mathrm{lb}} (|\tilde{\varrho}|Q -k) }{ k (|\tilde{\varrho}|Q/2)^2}},
\end{align*}
where $(a)$ is approximated because $({1+\varsigma^2})^{-\frac{1}{2}}\simeq (1-\frac{1}{2}\varsigma)$ for {a} sufficiently small $\varsigma$ based on the assumption $|x| \gg k $.
The ratio  {is greater} than one when the condition  ${\frac{\varrho_{q}^{\mathrm{ub}} \varrho_{q}^{\mathrm{lb}} (|\tilde{\varrho}|Q -k) }{ k (|\tilde{\varrho}|Q/2)^2}} > 1$  holds.
Note that this condition has been rewritten as $\varrho_{q}^{\mathrm{ub}} \varrho_{q}^{\mathrm{lb}}    >  {\frac{k (|\tilde{\varrho}|Q/2)^2}{|\tilde{\varrho}|Q -k}}$.
Assuming that the left-hand/right-hand edges of the sub-range are defined as a function of the position parameter, such that $\{\varrho_{q}^{\mathrm{lb}}, \varrho_{q}^{\mathrm{ub}} \} \doteq  \big\{ p -\frac{|\tilde{\varrho}|}{2}, p + \frac{|\tilde{\varrho}|}{2} \big\}$, the condition can be approximated as
\begin{align*}
|x| > \sqrt{\frac{|\tilde{\varrho}|^2}{4} + {\frac{k (|\tilde{\varrho}|Q/2)^2}{|\tilde{\varrho}|Q -k}} } \stackrel{(a)} \simeq  {\frac{\sqrt{|\varrho |(|\varrho |+kQ)}}{2}},
\end{align*}
where the approximation in $(a)$ holds with ${|\tilde{\varrho}|Q -k \simeq |\tilde{\varrho}|Q}$ because $|\varrho |Q \gg k$.
\end{IEEEproof}

Based on Parseval's theorem \cite{Ref_Hur13}, the total amount of energy with respect to the BP over the entire beam-region is fixed to the finite value $\frac{2\pi}{M}$.
In previous studies assuming a uniformly distributed beam direction, the length  of each beam-region is evenly divided to uniformly split the energy over the spatial frequency domain  \cite{Ref_Son14_2,Ref_Son17,Ref_Kim20}.
However, it is not reasonable to assume that the directions of a dominant radio path are uniformly distributed, because wireless coverage in a V2I network is very different from that in conventional cellular networks.
Assuming a uniform distribution of vehicles on the road, the beamforming gain must be uniformly distributed over the entire geographic range to maximize the average data-rate.

Around the center of the road, e.g., $|x|<k$, it is effective to equally divide the geographic sub-range.
However, as discussed in Lemma \ref{lm:01}, the beam width of the converted beam-region becomes  narrower than that of the equally divided beam-region, $|\tilde{\nu}|$, when the vehicle moves toward the road edge.
If the beam width of the converted beam-region is smaller than $\frac{2\pi}{M}$, {then the} total amount of energy, $\frac{2\pi}{M}$, cannot be contained within the beam-region because the maximum value of the normalized {BP} is limited to one.
For this reason, an oversampling effect occurs because neighboring codewords generate overlapped beam patterns.
As the vehicle drives to the edge in the geographical domain, the oversampling effect becomes severe because too many beamformers point in a similar direction, which is the end of the spatial frequency domain.
To compensate for the extreme oversampling effect, we  impose a minimum constraint on the beam width of a codeword.

\begin{figure}
\centering
\subfigure{\includegraphics[width=0.495\textwidth]{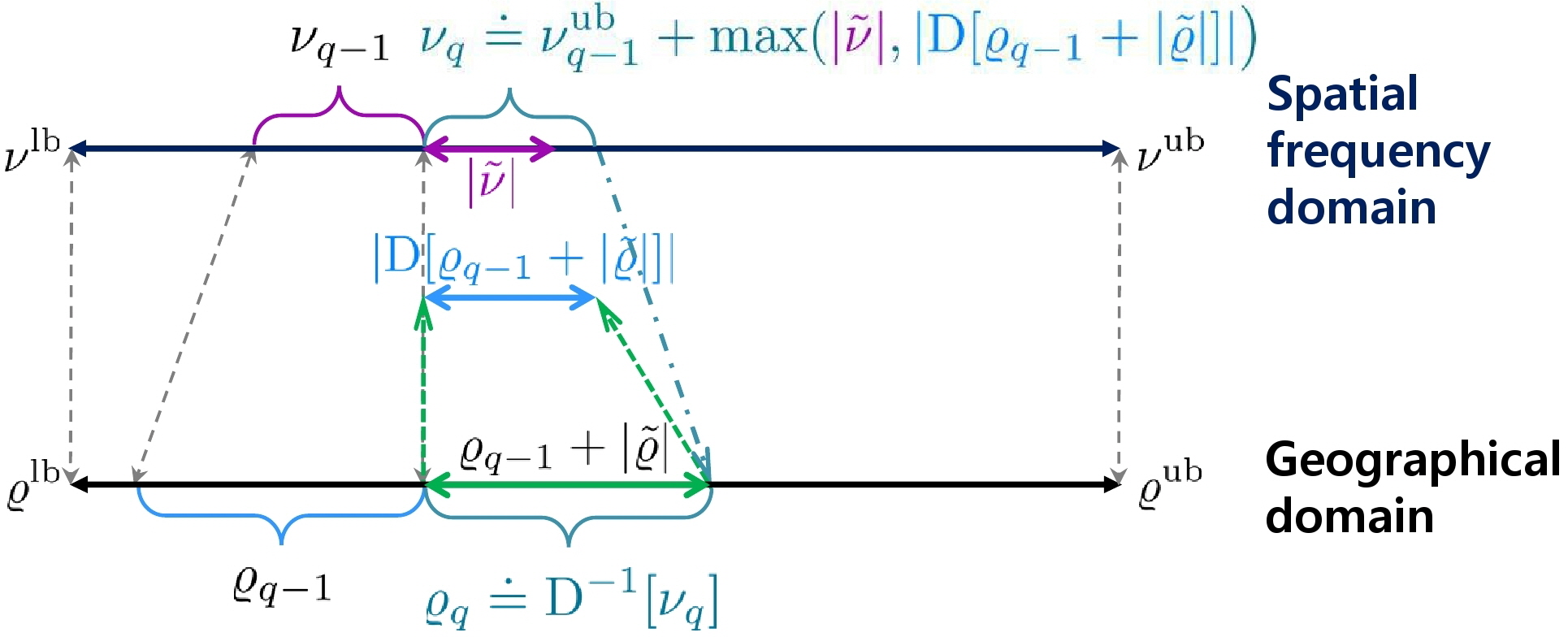}}
\caption{Proposed beam-region division algorithm.}
\label{fig:special_10}
\end{figure}

A beam-region division algorithm is proposed to resolve the overlapping BP problem.
First, the equally spaced geographic sub-range, $\varrho_{q-1}+|\tilde{\varrho}|$, {is} defined as the sub-range candidate.
The sub-range candidate is then converted into the spatial frequency domain, such that $\mathrm{D}[\varrho_{q-1}+|\tilde{\varrho}|]$, as depicted in Fig. \ref{fig:special_10}.
If the length of the converted beam-region is {greater} than that of the equally spaced beam-region {$|\tilde{\nu}|$}, {then the} sub-range candidate chosen will be  the $q$-th sub-range.
To avoid an overlapping BP  scenario, however, if the length of the converted beam-region is smaller than $|\tilde{\nu}|$, {then the} $q$-th beam-region will be designed to have {a} predefined minimum length $|\tilde{\nu}|$ in the spatial frequency domain.
Based on the beam-region division algorithm, the beam-region and geographic sub-range are defined as
$\big\{\nu_{q}, \varrho_{q} \big\} \doteq \big\{\nu_{q-1}^{\textrm{ub}}+\max\big(|\tilde{\nu}|,|\mathrm{D}[\varrho_{q-1}+|\tilde{\varrho}|]|\big), \mathrm{D}^{-1}[\nu_{q}] \big\}$.
If both the {geographical} and spatial frequency domains are initially divided by $Q$, such that $\{|\tilde{\nu}|, |\tilde{\varrho}|\} = \big\{ {\frac{2\nu^{\mathrm{ub}}}{Q}}, {\frac{2\varrho^{\mathrm{ub}}}{Q}} \big\}$, {then the} total number of {beam-regions that} are divided based on the proposed algorithm will be {less} than the predefined number of  codewords, $Q$.
For this reason, both domains will be  divided into more pieces than $Q$ until the set of $Q$ beam-regions is constructed.

\subsubsection{Codebook design for downlink beamforming}

\begin{figure}
\centering
\subfigure{\includegraphics[width=0.425\textwidth]{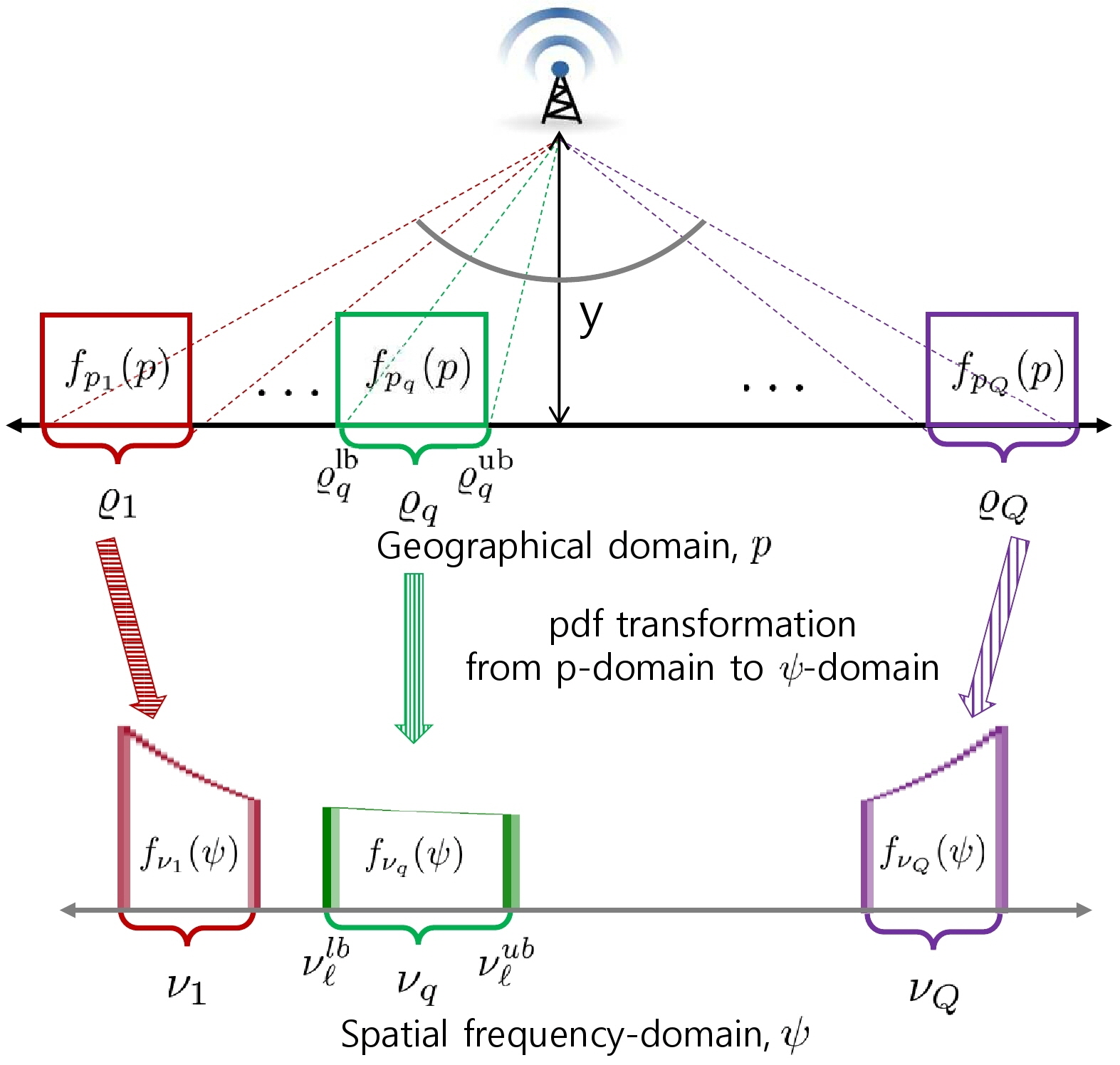}}
\caption{Domain transformation for beamforming codebook design.}
\label{fig:special_07}
\end{figure}

We design the transmit beamformer based on predefined geographic sub-ranges and beam-regions.
In the previous sub-section, the entire geographic range is divided into $Q$ non-overlapping  sub-ranges\footnote{After completing the beam-region division algorithm, the index restarts at 1 including negative beam-regions.}, such that $\{\varrho_{1},\cdots,\varrho_{Q} \}$.
In this paper, we focus on designing a transmit beamformer that generates a favorable BP in each target geographic range.

Assuming that the vehicle positions follow a uniform distribution along the road, the probability density function (pdf) of the position variable in $\varrho_{q}$ can be modeled by
\begin{align*}
f_{\varrho_{q}}(x)=\frac{\mathrm{u}(x-\varrho_{q}^{\mathrm{lb}})- \mathrm{u}(x-\varrho_{q}^{\mathrm{ub}})}{\varrho_{q}^{\mathrm{ub}}-\varrho_{q}^{\mathrm{lb}}}.
\end{align*}
{In the geographical domain, the BP should be uniformly distributed because the  data-rate is computed with respect to the uniformly distributed vehicle positions on the road.}

The position variable $x$ can be written as a function of spatial frequency  $\psi$, such that  $x=k\psi (\pi^2 - \psi^2)^{-\frac{1}{2}}$, as in (\ref{eq:spatial-geographic}).
Therefore, as depicted in Fig. \ref{fig:special_07}, the pdf  in the geographical {domain} can be transformed into the spatial frequency domain, such as
\begin{align}
\nonumber
f_{\nu_{q}}(\psi)&= f_{\varrho_{q}}\big(k\psi (\pi^2 - \psi^2)^{-\frac{1}{2}}\big)\bigg| \frac{\mathrm{d}}{ \mathrm{d}\psi}(k\psi (\pi^2 - \psi^2)^{-\frac{1}{2}})\bigg|
\\
\label{eq:psi_pdf}
\nonumber
&=\frac{\big|  \pi^2 k (\pi^2 - \psi^2)^{-3/2} \big|}{\varrho_{q}^{\mathrm{ub}}-\varrho_{q}^{\mathrm{lb}}}\Big[\mathrm{u}\big(k\psi (\pi^2 - \psi^2)^{-\frac{1}{2}}-\varrho_{q}^{\mathrm{lb}}\big)
\\
&~~~~~~~~~~~~~~~~~~~- \mathrm{u}\big(k\psi (\pi^2 - \psi^2)^{-\frac{1}{2}}-\varrho_{q}^{\mathrm{ub}}\big)\Big].
\end{align}

We design the beamforming codeword in order to generate a BP, which is similar to the pdf of the beam direction $f_{\nu_{q}}(\psi)$ in the spatial frequency domain.
The optimal BP of the $q$-th codeword is defined as
\begin{align}
\nonumber
\mathrm{G}^{\textrm{opt}}_{q}&(\psi)=\chi f_{\nu_{q}}(\psi)
 \\
\label{eq:ideal_BP}
\nonumber
& \stackrel{(a)} =  \frac{2\pi \big|  \pi^2 k (\pi^2 - \psi^2)^{-3/2} \big|}{M(\varrho_{q}^{\mathrm{ub}}-\varrho_{q}^{\mathrm{lb}})}\Big[\mathrm{u}\big(k\psi (\pi^2 - \psi^2)^{-\frac{1}{2}}-\varrho_{q}^{\mathrm{lb}}\big)
\\
&~~~~~~~~~~~~~~~~~~~~~-\mathrm{u}\big(k\psi (\pi^2 - \psi^2)^{-\frac{1}{2}}-\varrho_{q}^{\mathrm{ub}}\big)\Big],
\end{align}
where $\chi \in \mathbb{R}$ denotes the scaling parameter.
All BPs satisfy the integral constraint, $\int_{-\pi}^{\pi}\mathrm{G}^{\textrm{opt}}_{q} (\psi) \mathrm{d}\psi=\frac{2\pi}{M}$, based on Parseval’s theorem \cite{Ref_Hur13}.
The scaling coefficient in $(a)$ is thus computed as $\chi=\frac{2\pi}{M}$, as in \cite{Ref_Son14_2}.
Notably, beamforming gain is upper bounded by the unit gain.

We focus on minimizing the mean squared error between the optimal BP and actual BP to compute the beamforming codeword.
Similar to \cite{Ref_Son14_2,Ref_Son17}, we construct the beamformer, $\bu_{q}=\bF_{{q}}\bv_{{q}}$, such that
\begin{align}
\label{eq:cont_problem}
{(\bF_{q},\bv_{q}) = \argmin_{\tilde{\bF},\tilde{\bv}} \int_{-\pi}^{\pi} \Big| \mathrm{G}^{\textrm{opt}}_{q}(\psi) - \big|\bd_{M}^H(\psi) \tilde{\bF}\tilde{\bv}\big|^2  \Big|^2 \mathrm{d}\psi.}
\end{align}
The beamformer that minimizes the mean squared error in (\ref{eq:cont_problem}) is chosen {from} among the beamformer candidates, as in \cite{Ref_Son20BP}.
Please see  \cite{Ref_Son20BP} for more details {on} the method of generating beamformer {candidates, and for the} final codeword selection process.

\begin{figure}
\centering
\subfigure{\includegraphics[width=0.475\textwidth]{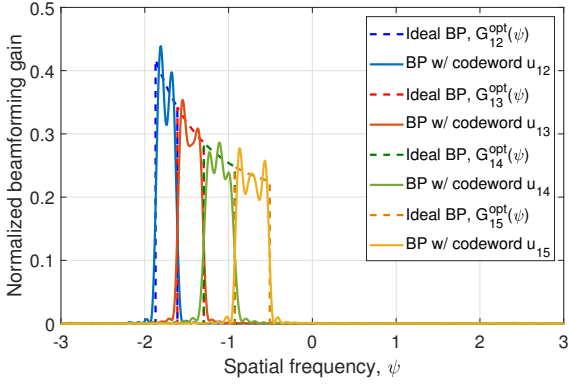}}
\caption{Comparison between ideal and actual BPs with $Q=32$ codewords in the beamforming codebook and $M=64$ transmit antennas.}
\label{fig:beam_pattern}
\end{figure}

\begin{figure}
\centering
\subfigure{\includegraphics[width=0.475\textwidth]{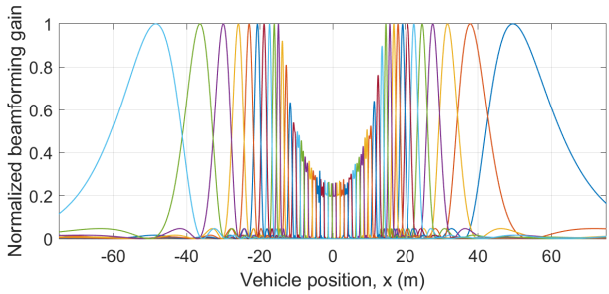}}
\caption{BP example of $Q=48$ codewords in the codebook $\mathcal{U}^{\textrm{BF}}_{48}$ and $M=96$ transmit antennas ($y=8.5$ m).}
\label{fig:beam_pattern_design}
\end{figure}

Finally, the beamforming codebook, which consists of $Q$ predefined codewords, is constructed as $\mathcal{U}^{\textrm{BF}}_{Q} \doteq \{\bu_{1},\cdots \bu_{Q} \}$.
Example BPs of beamforming codewords and codebook are depicted in {Figs. \ref{fig:beam_pattern} and \ref{fig:beam_pattern_design}}, respectively.
The dotted lines represent the ideal BP in {(\ref{eq:ideal_BP})}, and the solid lines represent the actual BP, generated by $\bu_{q}=\bF_{{q}}\bv_{{q}}$ in (\ref{eq:cont_problem}).
{{Fig. \ref{fig:beam_pattern} shows that} the beamformer design algorithm in \cite{Ref_Son20BP} is effective at generating a BP that is similar to the ideal BP in  {(\ref{eq:ideal_BP})}. }

\subsubsection{Beamformer selection algorithm}

The transmit beamformer should be chosen for a given set of state vectors $\hat{\bt}_{\tau}$ and covariance matrices $\bQ_{\tau}$ at time $\tau$.
Note that $\tau$ is the beamformer switching time, which is a nonnegative multiple of the channel sounding period  $T_s$.
Although uplink sounding is conducted for every $T_s$, it is not practical to recompute the transmit beamformer frequently.
The transmit beamformer, $\bc_{\tau}$, is periodically designed every $T_d=\Omega T_s$ seconds.
The beamformer $\bc_{\tau}$ at time instance $\tau$ should be chosen to provide mobile service with a large beamforming gain  within the next period, i.e., $\{ (\tau+1),\cdots, (\tau+ \Omega)\}$.

First, we  estimate the future state vectors without {the} aid of  uplink sounding.
The estimated state vectors are denoted by $\{\tilde{\bt}_{\tau+1},\cdots, \tilde{\bt}_{\tau + \Omega}\}$.
Without the state update process based on the uplink sounding observations, the predicted state vector and covariance matrix cannot be corrected using the Kalman gain matrix in (\ref{eq:kalman}).
Therefore, the transmit beamformer for the next $T_d$ seconds has to be designed using the state vectors and covariance matrices that are estimated based only on the state prediction process.
For a given estimated state vector $\hat{\bt}_{\tau}$ at time instance $\tau$, the state vectors are predicted as
\begin{align*}
\tilde{\bt}_{\tau+t}&=\bA^{t}\hat{\bt}_{\tau},~~\textrm{w/o estimated}~\hat{\alpha},
\\
\tilde{\bt}_{\tau+t}&=\bA^{t}\hat{\bt}_{\tau}+{\bb}\hat{\alpha},~~\textrm{w/ estimated}~\hat{\alpha}.
\end{align*}
Furthermore, the predicted covariance matrix is predicted as
\begin{align*}
\tilde{\bQ}_{\tau+t}&=\bA^{t}{\bQ}_{\tau}(\bA^T)^{t}+{\sum}_{\tau=0}^{t-1}\bA^{\tau}(\bQ_{\alpha}+\bQ_{\omega})(\bA^T)^{\tau},~\textrm{w/o}~\hat{\alpha},
\\
\tilde{\bQ}_{\tau+t}&=\bA^{t}{\bQ}_{\tau}(\bA^T)^{t}+{\sum}_{\tau=0}^{t-1}\bA^{\tau}\bQ_{\omega}(\bA^T)^{\tau},~\textrm{w/}~\hat{\alpha}.
\end{align*}

The statistical distribution of the position variable is defined for the given estimated state vector and covariance matrix.
In the estimated state vector, the position variable on the $x$-axis can be modeled by $x_{\tau+t} =\tilde{x}_{\tau+t}+\nu_{\tau+t}$, where $\tilde{x}_{\tau+t}=(\tilde{\bt}_{\tau+t})_{1,1}$, $\nu_{\tau+t} \sim \mathcal{N}(0,\sigma_{\tau+t}^2)$, and  $\sigma_{\tau+t}^2=(\tilde{\bQ}_{\tau+t})_{1,1}$.
The position variable on the $y$-axis is defined by  ${y}_{\tau+t}=(\tilde{\bt}_{\tau+t})_{2,1}$.
Furthermore, the beam direction  {at  position} $x_{\tau+t}$ can be approximated by
\begin{align*}
\psi_{\tau+t} &\simeq q(\tilde{x}_{\tau+t},{y}_{\tau+t}) +\frac{\partial q(x,{y}_{\tau+t})}{\partial x}\bigg|_{x=\tilde{x}_{\tau+t}}(x_{\tau+t}-\tilde{x}_{\tau+t})
\\
&=q(\tilde{x}_{\tau+t},{y}_{\tau+t})+ \dot{q}(\tilde{x}_{\tau+t},{y}_{\tau+t})(x_{\tau+t}-\tilde{x}_{\tau+t}),
\end{align*}
where $q(\tilde{x}_{\tau+t},{y}_{\tau+t}) \doteq {\pi \tilde{x}_{\tau+t}}{(\tilde{x}_{\tau+t}^2+{y}_{\tau+t}^2+h^2)^{-\frac{1}{2}}}$, and $\dot{q}(\tilde{x}_{\tau+t},{y}_{\tau+t}) ={\pi ({y}_{\tau+t}^2+h^2)}{(\tilde{x}_{\tau+t}^2+{y}_{\tau+t}^2+h^2)^{-\frac{3}{2}}}$.
The beam direction is then modeled {as a} normal distribution,   $\psi_{\tau+t} \sim \mathcal{N}\big(q(\tilde{x}_{\tau+t},{y}_{\tau+t}),\dot{q}(\tilde{x}_{\tau+t},{y}_{\tau+t})^2\sigma_{\tau+t}^2\big)$.

\begin{figure*}[!t]
\normalsize
\centering
\subfigure[Position estimation ($x$-axis)]{\label{fig:tracked_result_a}\includegraphics[width=0.425\textwidth]{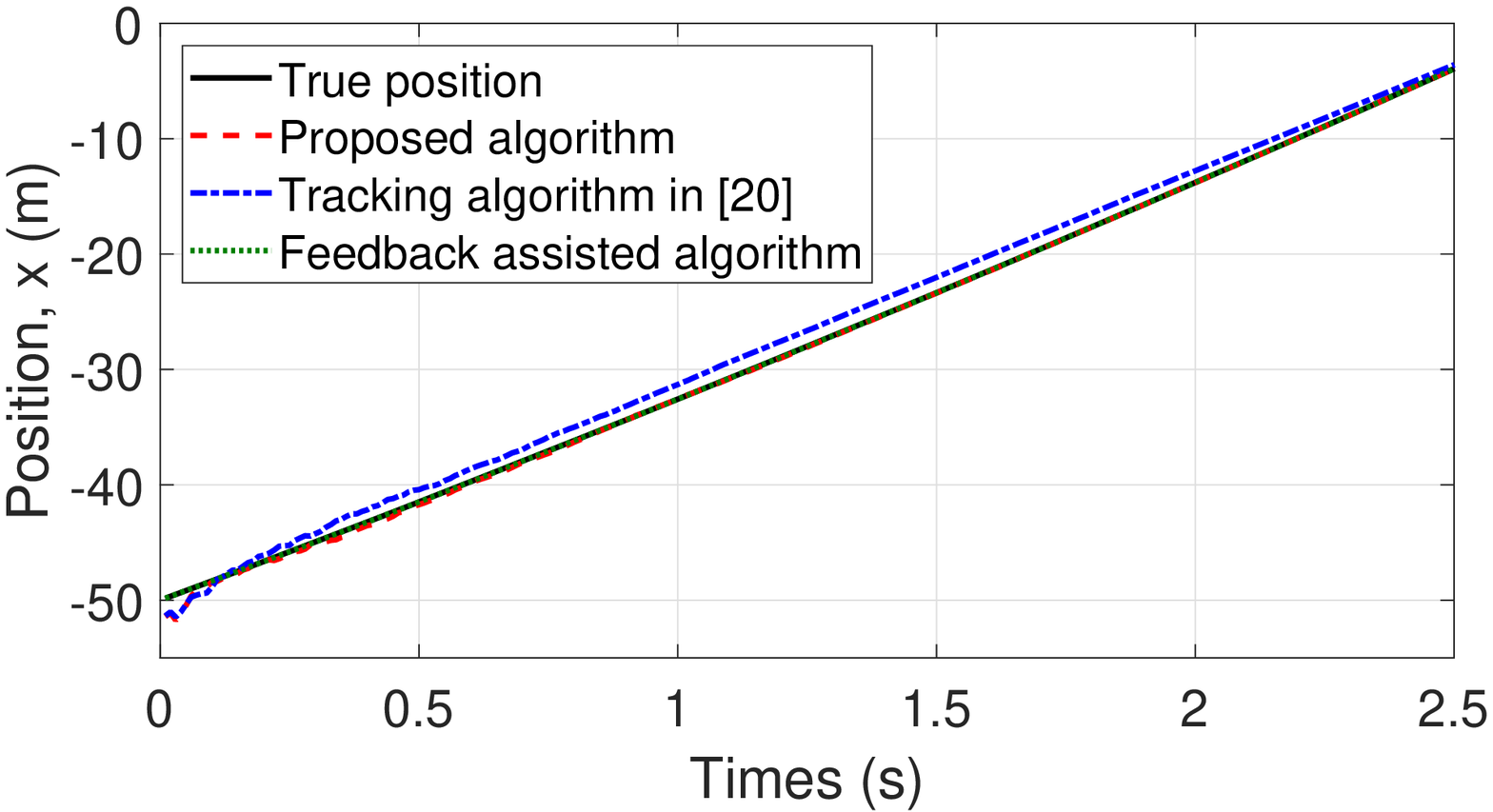}}
\hfil
\subfigure[Position estimation ($y$-axis)]{\label{fig:tracked_result_b}\includegraphics[width=0.425\textwidth]{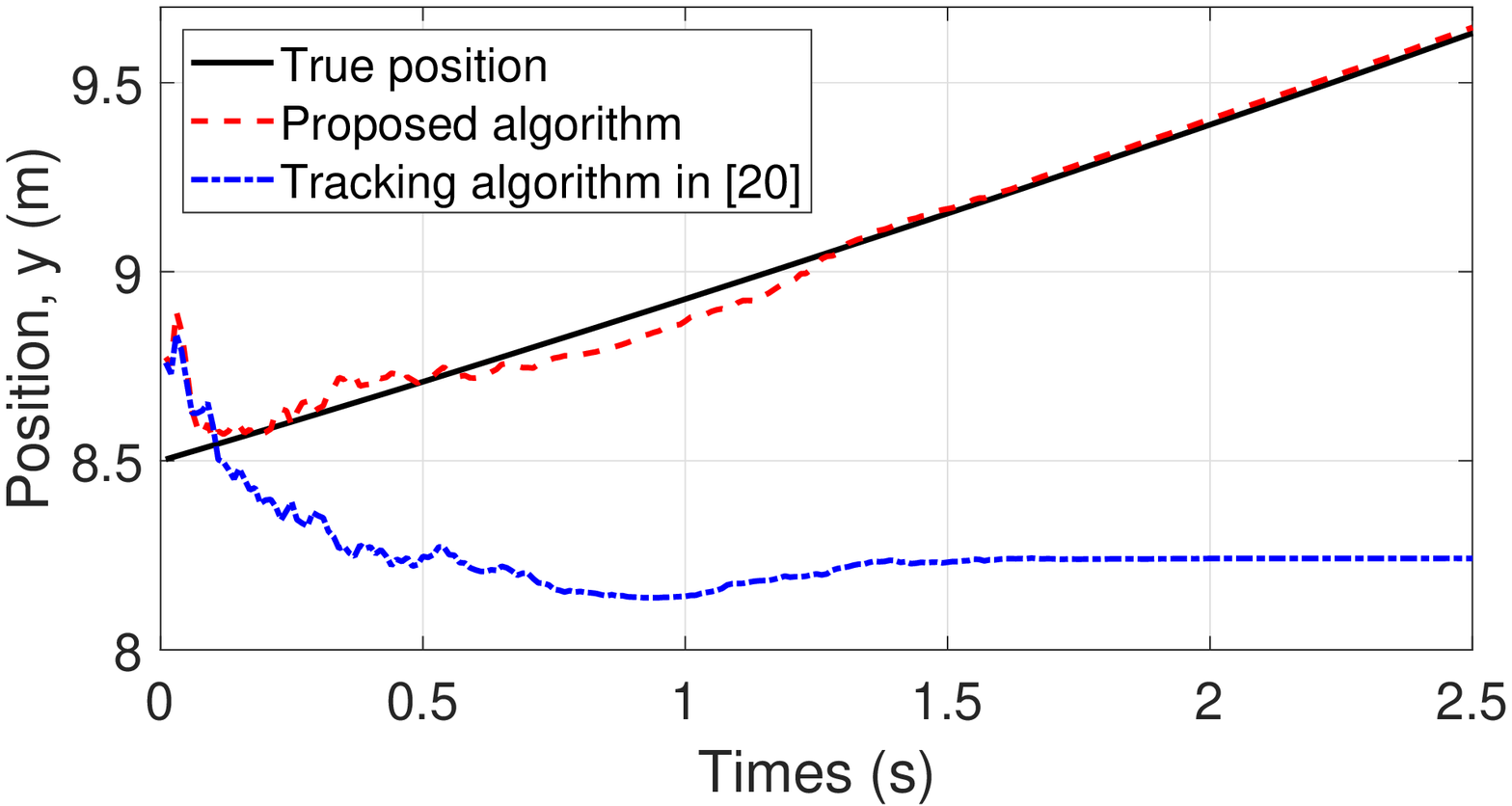}}
\hfil
\subfigure[Velocity estimation]{\label{fig:tracked_result_c}\includegraphics[width=0.425\textwidth]{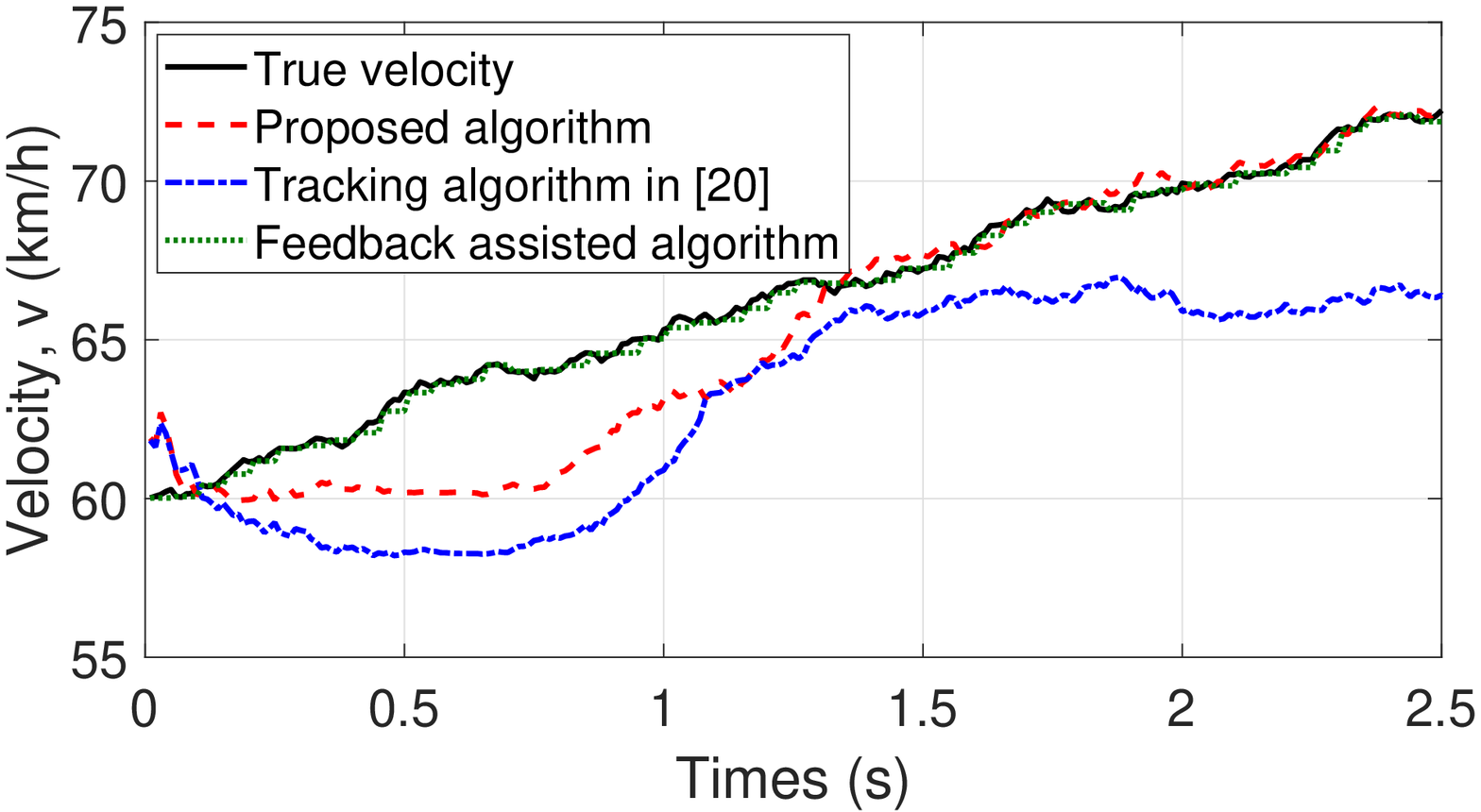}}
\hfil
\subfigure[Acceleration estimation]{\label{fig:tracked_result_d}\includegraphics[width=0.425\textwidth]{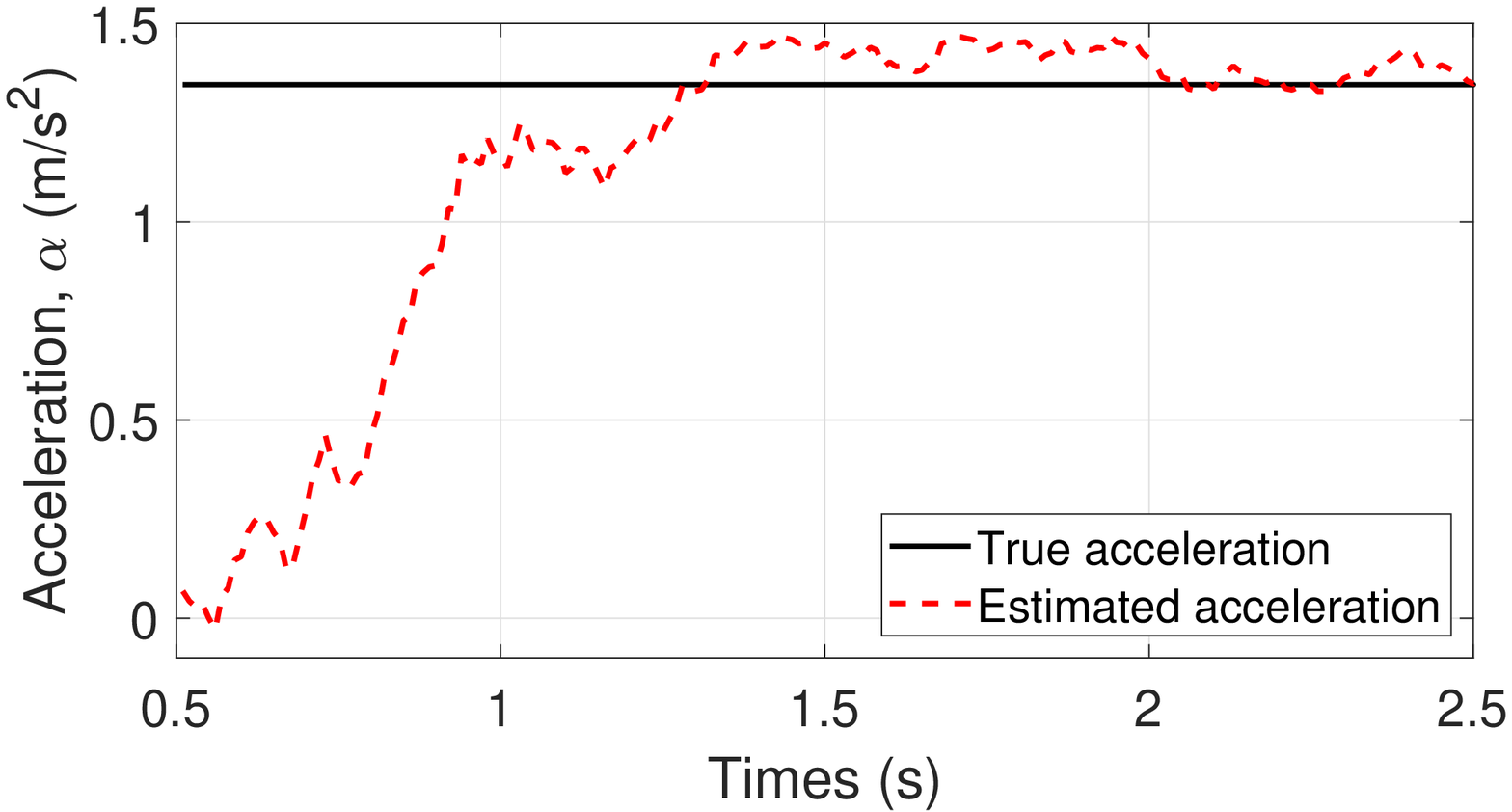}}
\caption{{Examples of vehicle tracking and acceleration estimation algorithms with $M=128$, $\varrho=-20$ dB, and $\sigma_{\epsilon}=10^{-1.5}$.}}
\label{fig:tracked_result}
\end{figure*}

Second, we define an ideal BP based on the statistical distribution of the estimated beam directions $\{\psi_{\tau+1},\cdots,\psi_{\tau+\Omega}\}$.
We  define the BP that maximizes the average beamforming gain within the next $T_d$ seconds.
The ideal BP is defined by solving the following problem:
\begin{align*}
&\mathrm{G}_{\tau}^{\textrm{ideal}}(\psi)
=\argmax_{\bar{\mathrm{G}}({\psi})}  \int_{-\pi}^{\pi} \bar{\mathrm{G}}({\psi}) \bigg[ \sum_{t=1}^{\Omega} f_{\psi_{\tau+t}}(\psi)\bigg] \mathrm{d}\psi
\\
& \stackrel{(a)}  \leq  \bigg(\int_{-\pi}^{\pi} \bar{\mathrm{G}}({\psi})^2   \mathrm{d}\psi\bigg)^{-\frac{1}{2}} \bigg( \int_{-\pi}^{\pi}  \bigg[ \sum_{t=1}^{\Omega} f_{\psi_{\tau+t}}(\psi)\bigg]^2 \mathrm{d}\psi \bigg)^{-\frac{1}{2}},
\end{align*}
where $(a)$ is derived based on the {Cauchy-Schwarz-{Buniakovsky}} inequality.
The equality in $(a)$ holds when the ideal BP is defined by
$\mathrm{G}_{\tau}^{\textrm{ideal}}(\psi)=\chi\sum_{t=1}^{\Omega} f_{\psi_{\tau+t}}(\psi)$.
Since BP {satisfies} the constraint, $\int_{-\pi}^{\pi}\mathrm{G}_{\tau}^{\textrm{ideal}}(\psi) \mathrm{d}\psi=\frac{2\pi}{M}$, as in \cite{Ref_Hur13}, the scaling parameter is computed based on the following equation,
\begin{align*}
\int_{-\pi}^{\pi}\mathrm{G}_{\tau}^{\textrm{ideal}}(\psi) \mathrm{d}\psi&=\chi \sum_{t=1}^{\Omega} \int_{-\pi}^{\pi}f_{\psi_{\tau+t}}(\psi)  \mathrm{d}\psi \stackrel{(a)} \simeq \chi {\Omega}   =\frac{2\pi}{M},
\end{align*}
where $(a)$ holds because $\int_{-\pi}^{\pi}f_{\psi_{\tau+t}}(\psi) \mathrm{d}\psi \simeq 1$.
The ideal BP is defined by
\begin{align}
\label{eq:pdfs}
\mathrm{G}_{\tau}^{\textrm{ideal}}(\psi)=\frac{2\pi}{M{\Omega}}\sum_{t=1}^{\Omega} f_{\psi_{\tau+t}}(\psi).
\end{align}

Finally, we propose a method for choosing the beamformer from the beamforming codebook $\mathcal{U}^{\textrm{BF}}_{Q}$.
We choose a transmit beamformer that can generate a  BP  similar to the ideal BP in  (\ref{eq:pdfs}).
To effectively compare the beamformer's BP with the ideal BP, each codeword $\bu_{q}$ is mapped {to} the actual BP vector, $\bg^{\textrm{actual}}_{q}  \in \mathbb{R}^L$, consisting of a finite number of beamforming gain samples.
Furthermore, for a given $\mathrm{G}_{\tau}^{\textrm{ideal}}(\psi)$, we define an ideal BP vector, $\bg^{\textrm{ideal}}_{\tau} \in \mathbb{R}^{L}$, consisting of $L$ {samples} in  $\mathrm{G}_{\tau}^{\textrm{ideal}}(\psi)$.
We compare the ideal BP vector, $\bg^{\textrm{ideal}}_{\tau}$, with the actual BP vector $\bg^{\textrm{actual}}_{q}$.
The {selected transmit beamformer} is $\bc=\bu_{\hat{q}}$, where the selected codeword index is
\begin{align*}
\hat{q} = \argmax_{q \in \{1,\cdots,Q\}} \big| (\bg^{\textrm{ideal}}_{\tau})^T \bg^{\textrm{actual}}_{q} \big|^2.
\end{align*}

\section{Simulation Results}
\label{sec:numerical}

We provide numerical results {by} evaluating the proposed vehicle tracking algorithm and beamforming codebook design.
We consider a hybrid beamforming system using $M=64, 96,$ and $128$ transmit antenna elements that are controlled by $N=4$ radio frequency  chains.
For Monte-Carlo simulations, we generated $10,000$ independent channels consisting of a line-of-sight and a non-line-of-sight radio path with Rician {$K$-factor of $13$ dB}.
We consider a system with 20 MHz bandwidth with a center frequency of $f_c=28$ GHz.
The noise power integrated over a given bandwidth is $\sigma_{\mathrm{n}}^2=-174+10\log_{10}(20 \times 10^6)\simeq -101$ dBm and the path-loss exponent is set to $n=2$.
The uplink sounding  period is  $T_s=10$ ms, and the downlink transmission period is  $T_d=\Omega T_s$ ms, with $\Omega \in \{10, 20\}$.

\begin{figure*}[!t]
\normalsize
\centering
\subfigure[Position tracking performance ($x$-axis).]{\label{fig:sim_re_00_01}\includegraphics[width=0.45\textwidth]{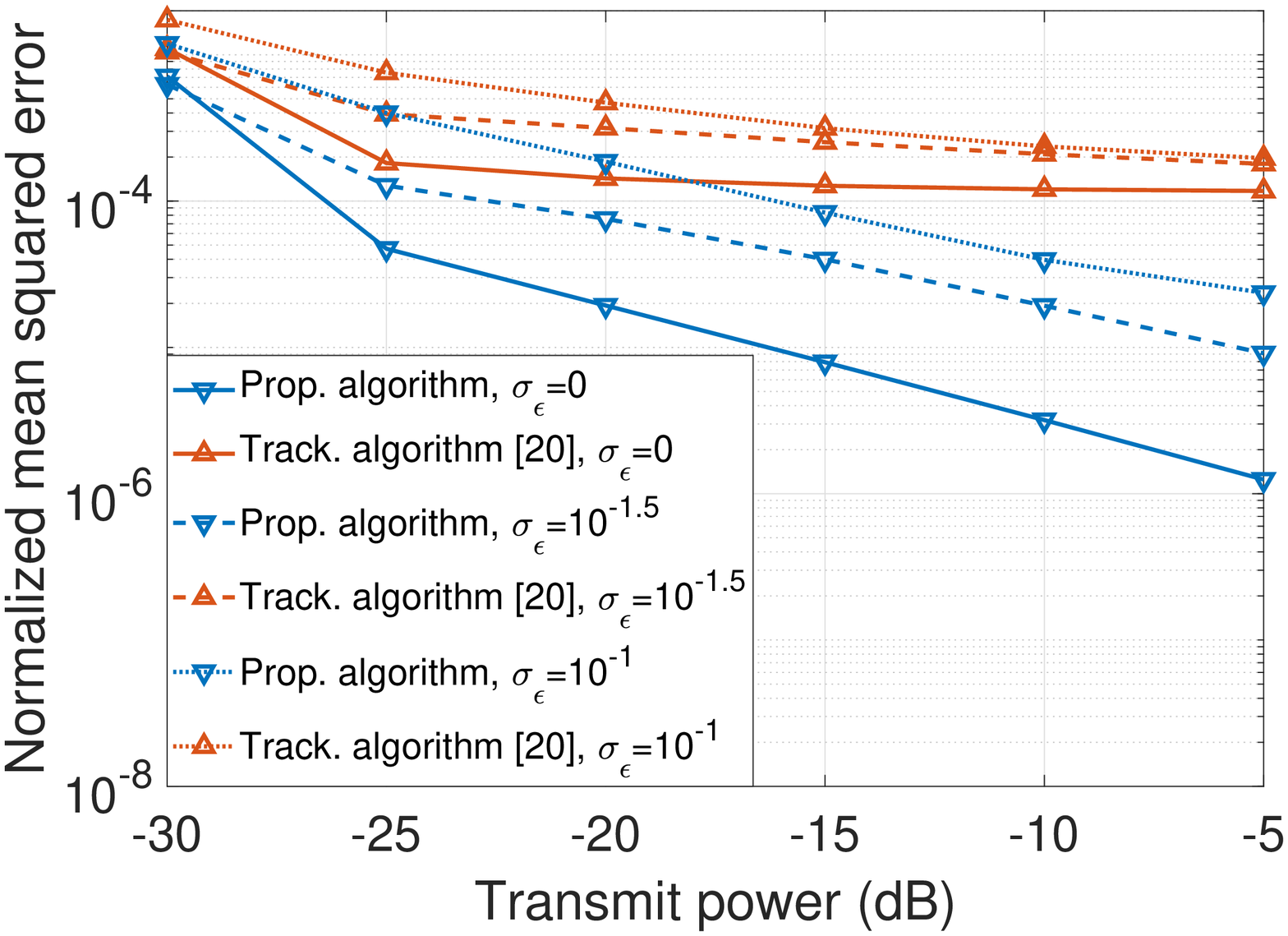}}
\hfil
\subfigure[Velocity tracking performance.]{\label{fig:sim_re_00_02}\includegraphics[width=0.45\textwidth]{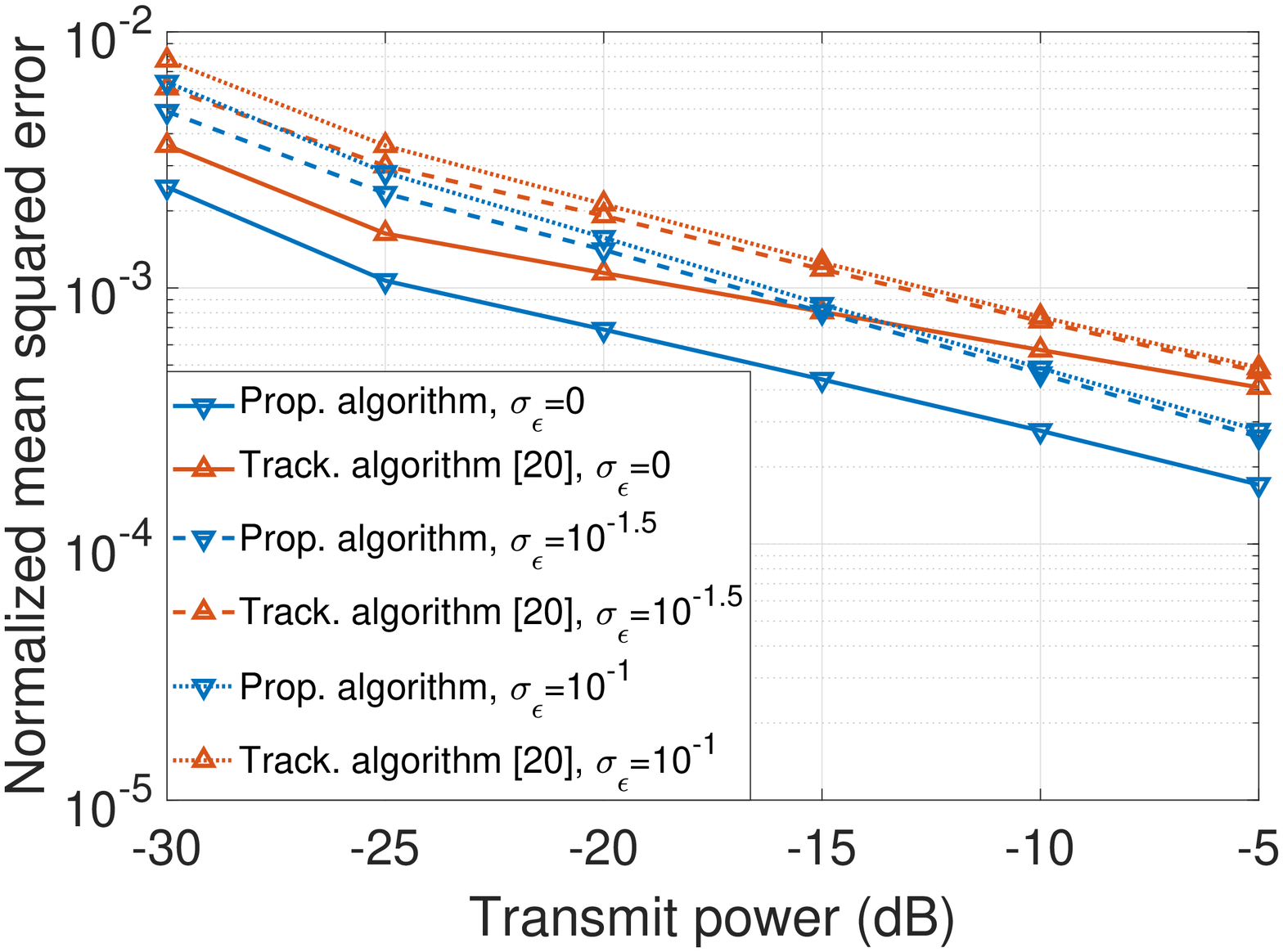}}
\caption{{Normalized mean squared error of vehicle tracking algorithms with $M = 96$ and $\varphi=\pi/2^{9}$.}}
\label{fig:sim_result_00}
\normalsize
\centering
\subfigure[Normalized beamforming gain $M=64$.]{\label{fig:sim_re_01}\includegraphics[width=0.425\textwidth]{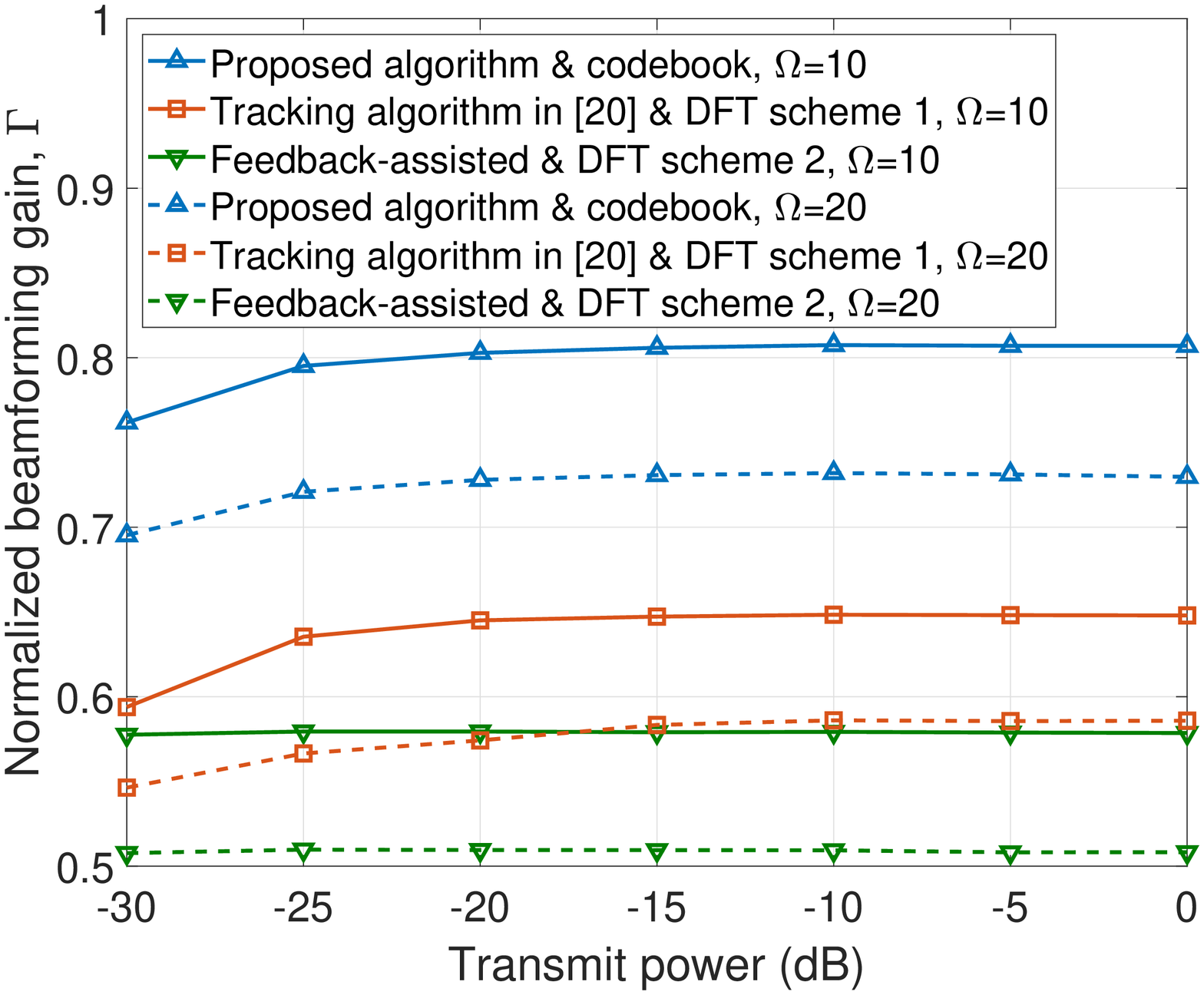}}
\hfil
\subfigure[Normalized beamforming gain $M=96$.]{\label{fig:sim_re_02}\includegraphics[width=0.425\textwidth]{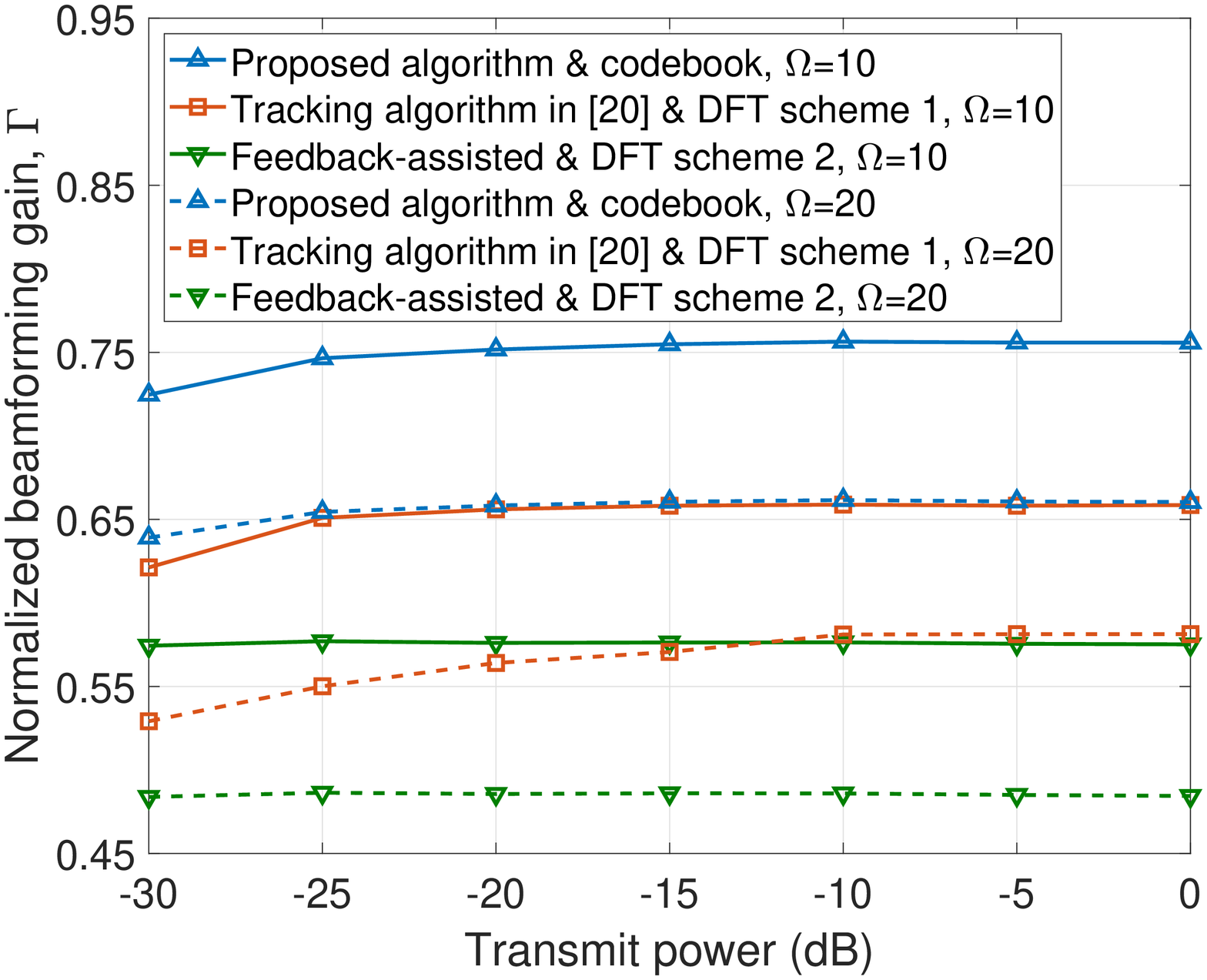}}
\hfil
\subfigure[Data-rate $M=64$.]{\label{fig:sim_re_01}\includegraphics[width=0.425\textwidth]{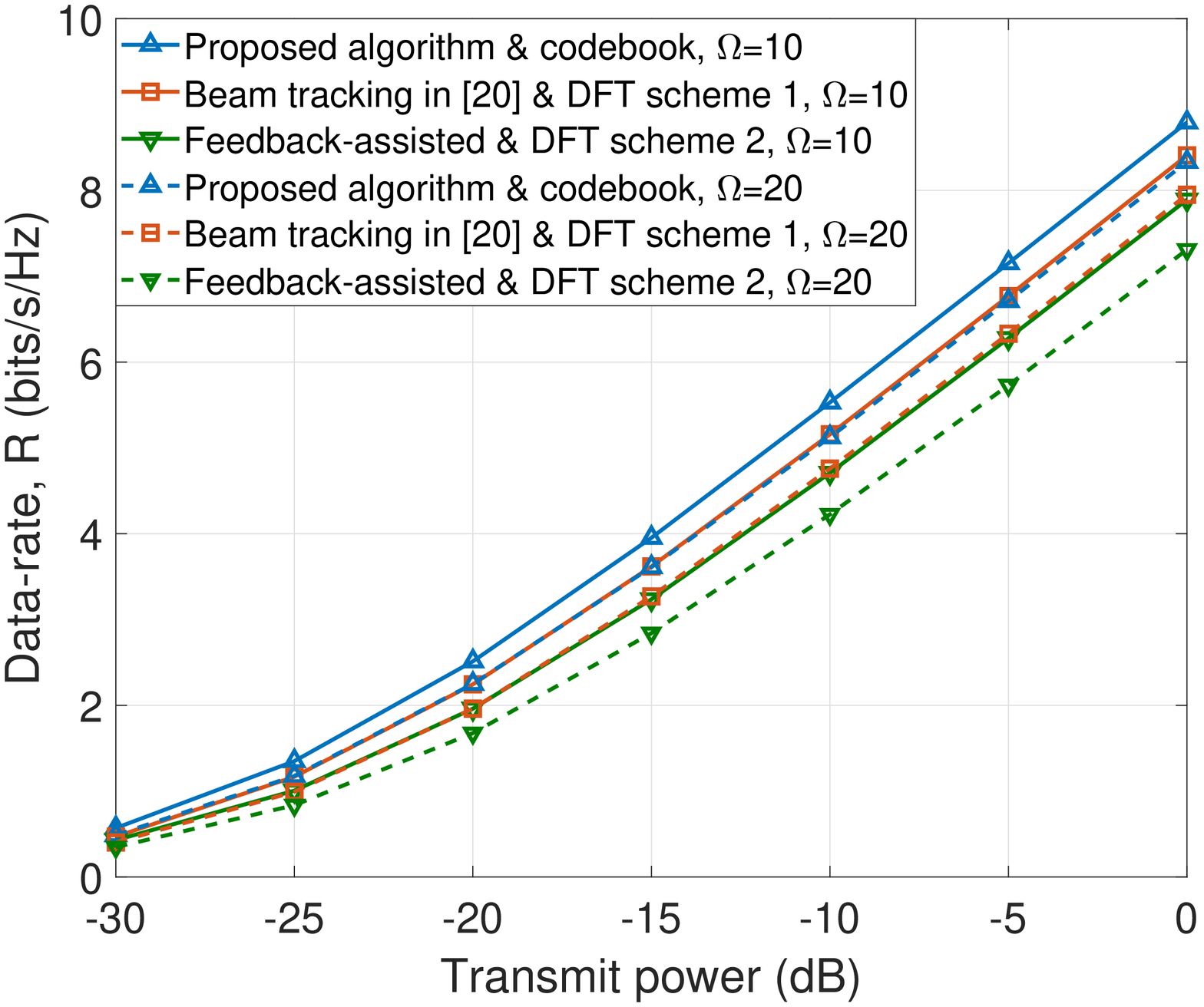}}
\hfil
\subfigure[Data-rate $M=96$.]{\label{fig:sim_re_02}\includegraphics[width=0.425\textwidth]{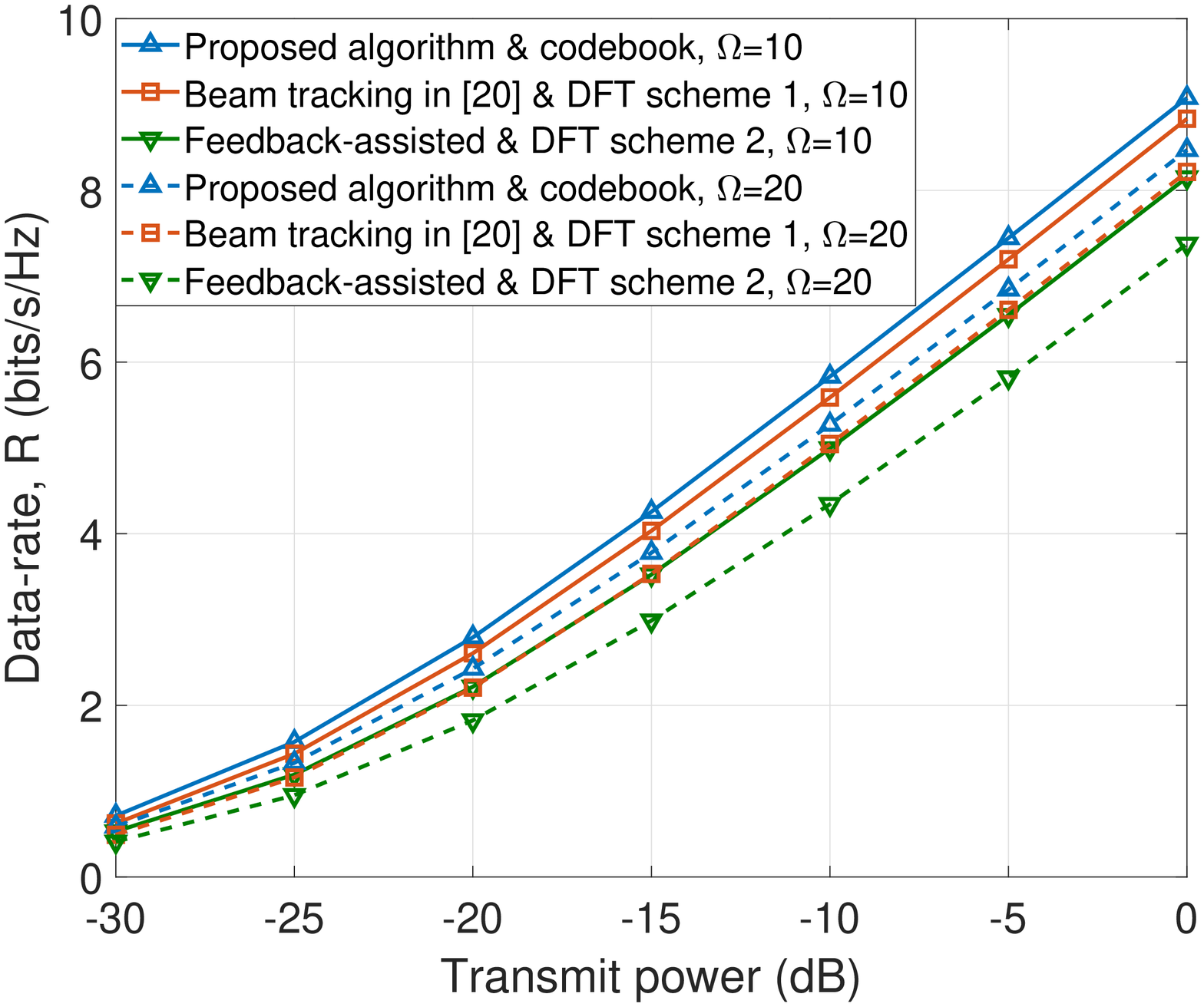}}
\caption{{Beamforming gain and data-rate of single-resolution beamforming codebook $(\Omega=10, 20,~Q=Q_1=M,~O=1,~\sigma_{\epsilon}=0 )$.}}
\label{fig:sim_result_01}
\end{figure*}

\begin{figure*}[!t]
\normalsize
\centering
\subfigure[Normalized beamforming gain $M=64$.]{\label{fig:sim_re_01}\includegraphics[width=0.4625\textwidth]{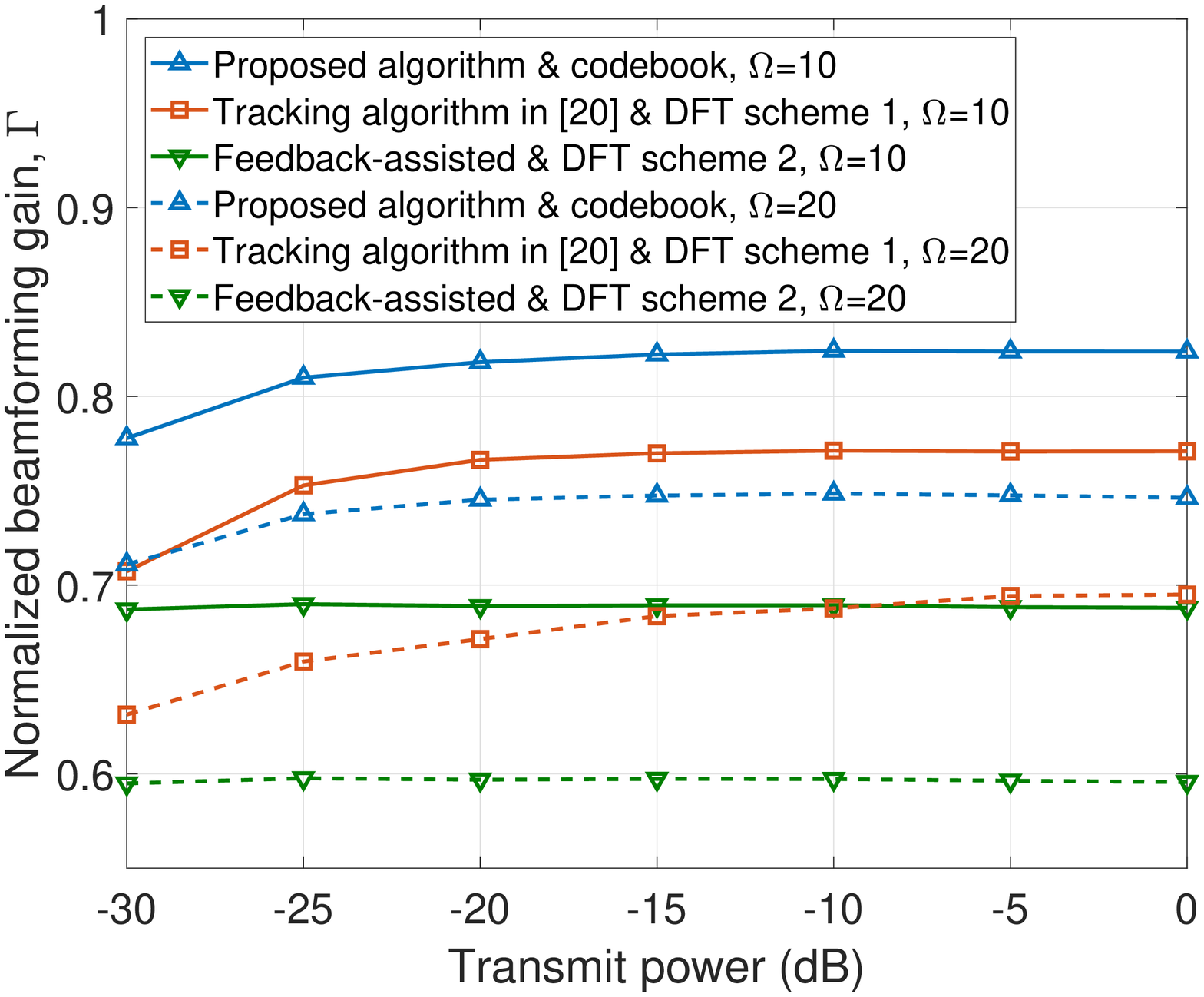}}
\hfil
\subfigure[Normalized beamforming gain $M=96$.]{\label{fig:sim_re_02}\includegraphics[width=0.4625\textwidth]{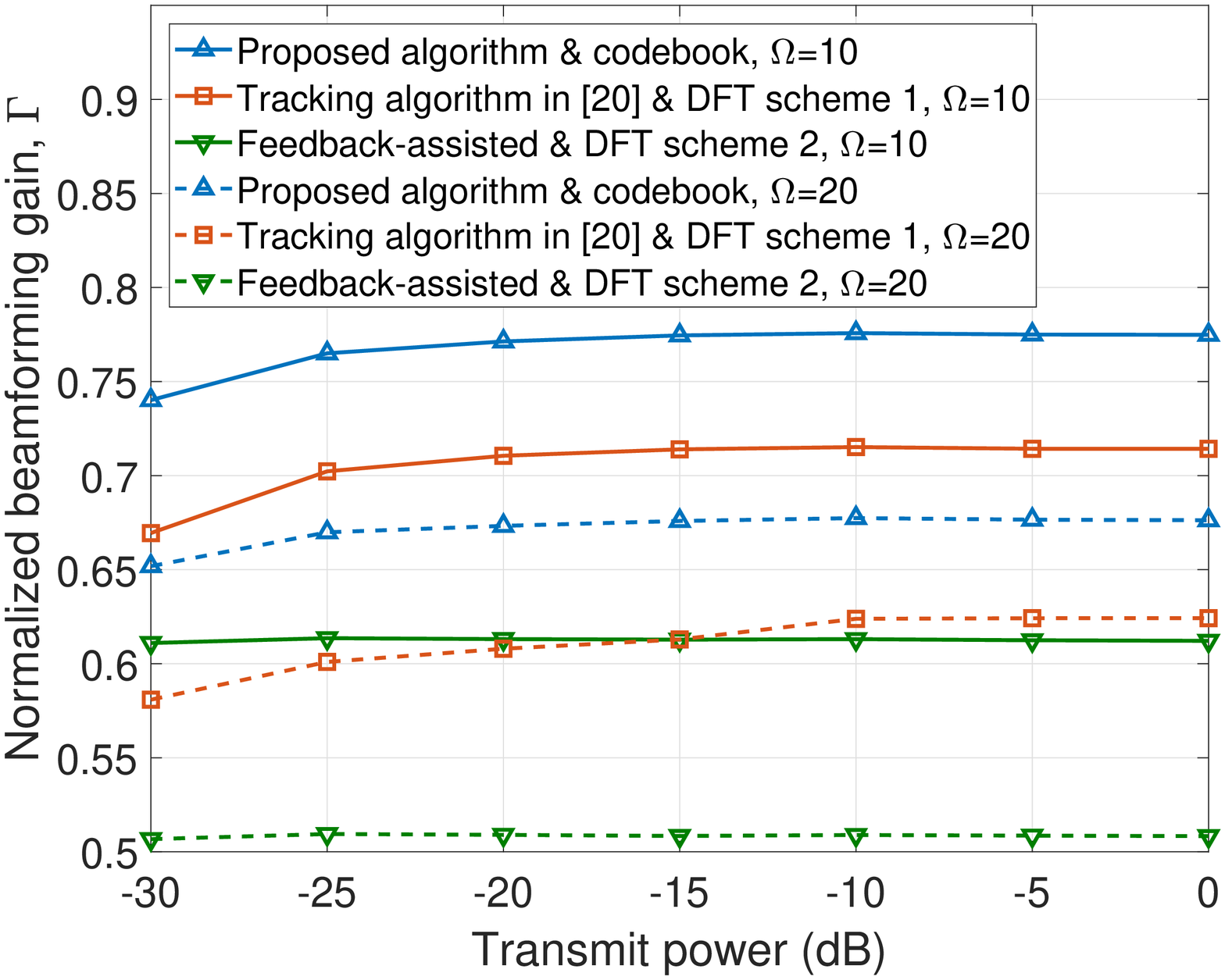}}
\hfil
\subfigure[Data-rate $M=64$.]{\label{fig:sim_re_01}\includegraphics[width=0.4625\textwidth]{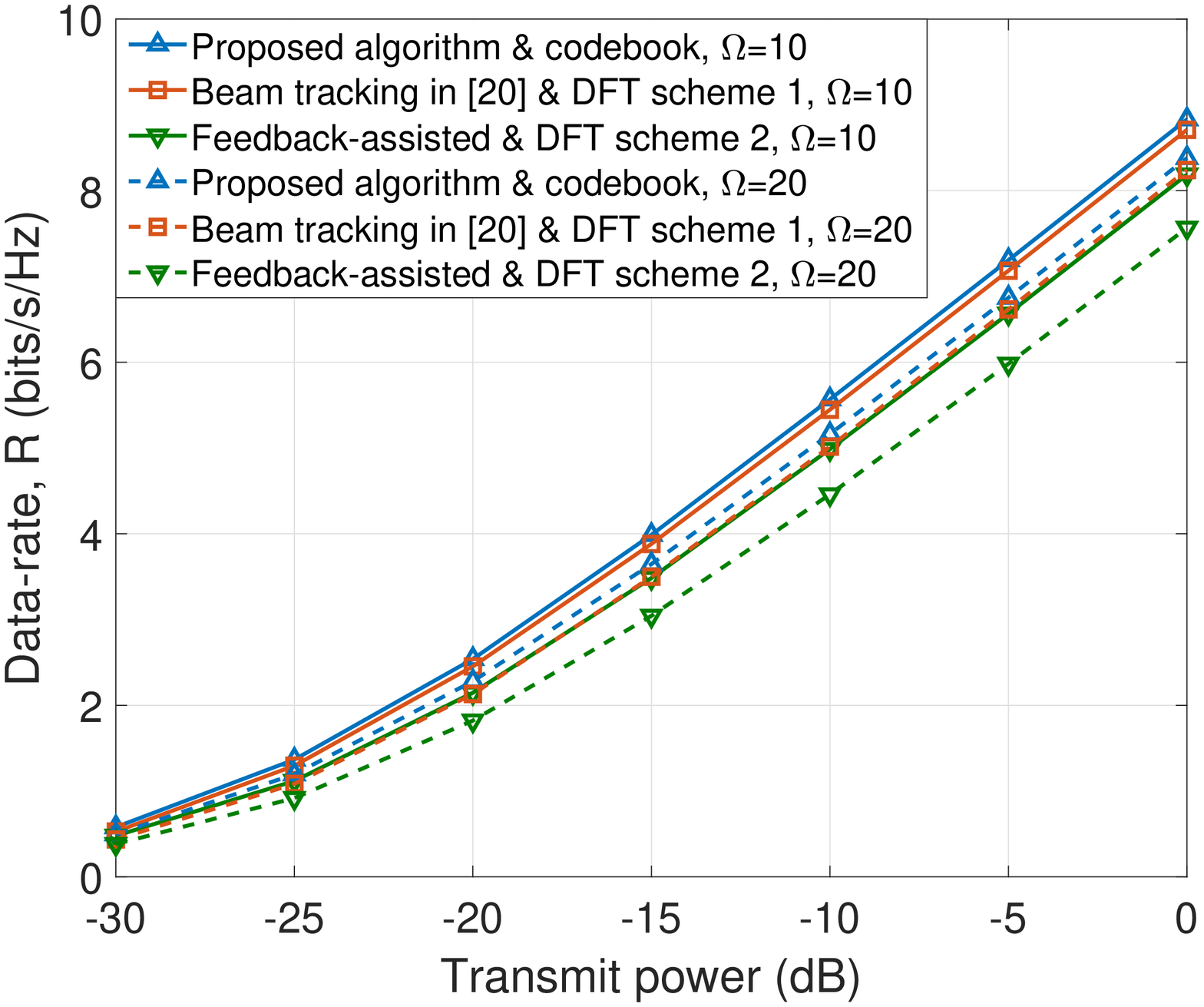}}
\hfil
\subfigure[Data-rate $M=96$.]{\label{fig:sim_re_02}\includegraphics[width=0.4625\textwidth]{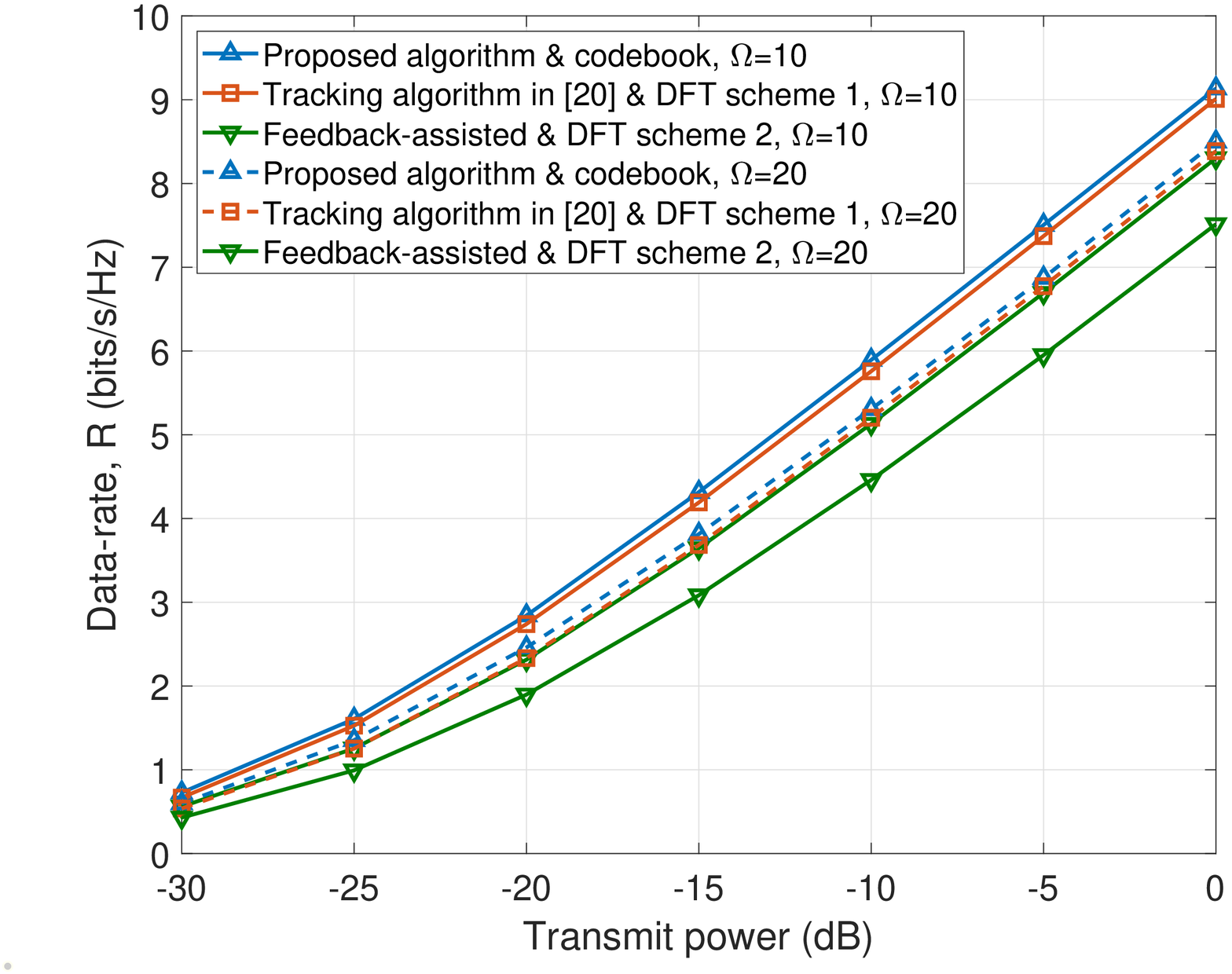}}
\caption{{Beamforming gain and data-rate of multi-resolution beamforming codebook $(\Omega=10, 20,~(Q_1,Q_2)=(\frac{M}{2},M),~Q=\frac{3M}{2},~O=2,~\sigma_{\epsilon}=0)$.}}
\label{fig:sim_result_02}
\end{figure*}

The height of RSU is $h=7.5$ m, and the vehicle steering angle is $\varphi=\pi/2^{7}$.
The left-hand edge/right-hand edges of the entire geographic range are defined as $(\varrho^{lb},\varrho^{ub})=(-75~\textrm{m},75~\textrm{m})$.
In the linear state prediction model, the acceleration parameter is assumed to follow the Gaussian distribution $\alpha \sim \mathcal{N}(0,\sigma_{\alpha}^2)$ with standard deviation $\sigma_{\alpha}=10^{-1}\big(v_0\frac{ 10^3}{60^2}\big)$ and  initial velocity parameter $v_0$.
Furthermore, the error parameter is assumed to follow the Gaussian distribution with standard deviation $\sigma_{\omega}=10^{-1.5}$.
Finally, the threshold for the acceleration estimation process is set to  $\alpha_{\mathrm{thres}}=3\%$.

We evaluate the  estimation performance of the proposed vehicle tracking algorithm.
We provide examples of state estimation results. For the numerical analysis, the initial state vector is set to ${\bt}_{0}=[-50~\textrm{m}, 8.5~\textrm{m}, 70~\textrm{km/h}]^T$.
To reflect the initialization error, the feedback error parameter is assumed to follow the Gaussian distribution with standard deviation $\sigma_{\epsilon}= 10^{-1.5}$.
The state estimation results of the proposed {vehicle tracking} algorithm are compared with those of  the {vehicle tracking} algorithm\footnote{{The {vehicle tracking} algorithm in \cite{Ref_Sha19} is modified to track the vehicle movement on the $x$-axis by considering the V2I transmission framework depicted in Fig. \ref{fig:system_overview}.}} in \cite{Ref_Sha19} and those of the feedback-assisted state tracking approach.
In the feedback-assisted state tracking method, the actual state vector is fed back to RSU periodically every $5T_s$, via a feedback link.

As examined in {Figs.} \ref{fig:tracked_result_a} and \ref{fig:tracked_result_b}, the position variables  are well estimated because equation (\ref{eq:linearized_io}), which is  used in the vehicle tracking process, is defined as a function of the vehicle position.
On the other hand, the velocity estimation illustrated in Fig. \ref{fig:tracked_result_c} is relatively worse than the estimation of the position variable because the velocity variable is not directly connected to equation (\ref{eq:linearized_io}) for the vehicle tracking process.
Furthermore, due to the initial feedback error, there is a difference between the estimated state vector and true state vector in the initial stage of the tracking process.
However, with the uplink channel sounding from the vehicle, the effect of the feedback error can be properly corrected.
Hence, the effect of the initial feedback error gradually diminishes as the EKF-based vehicle tracking algorithm progresses.
As shown in Figs. \ref{fig:tracked_result_a} and \ref{fig:tracked_result_c}, we performed a comparison with the proposed EKF-based vehicle tracking algorithm and feedback-based tracking method.
In the case of the feedback-based tracking algorithm, the vehicle directly transmits position and velocity information to RSU. Therefore, it can provide a higher estimation performance than the EKF-based algorithm.
Unlike the EKF algorithm which relies upon sounding signals, the feedback-based algorithm requires additional data transfer fields, which may dramatically increase the payload overhead for the feedback process.
Fig. \ref{fig:tracked_result_c}  shows that the difference between the actual and estimated velocities decreases over time.
These results can be explained by the changes in the acceleration estimation performance depicted in Fig. \ref{fig:tracked_result_d}.
After the proposed acceleration algorithm converges, the estimated acceleration parameter is applied to the state prediction process.
Consequently, the position and velocity estimation performance are improved as the acceleration parameter is accounted for, and RSU can solely focus on correcting the error parameter.
The state estimation results show that the acceleration estimation will get better as the vehicle tracking performance is enhanced, and vice versa.
The vehicle tracking results should be accumulated in order to obtain an estimated acceleration parameter that can be applied to the vehicle tracking process.

In Figs. \ref{fig:sim_re_00_01} and \ref{fig:sim_re_00_02}, {the} position and velocity tracking performances are evaluated using {the} normalized mean squared error (NMSE), $\Upsilon_{x}=\mathrm{E}\big[|\frac{x_{\ell}-\hat{x}_{\ell}}{x_{\ell}}|^2\big]$ and $\Upsilon_{v}=\mathrm{E}\big[|\frac{v_{\ell}-\hat{v}_{\ell}}{v_{\ell}}|^2\big]$, respectively.
To check the effect of the initial feedback error, the standard deviation {of the feedback error} is set to $\sigma_{\epsilon}=[0,10^{-1.5},10^{-1}]$.
The simulation results show that a smaller NMSE can be achieved when the initial feedback error has less impact on the state vector.
It is also shown that the NMSE decreases with increasing SNR, which implies that RSU can obtain measurement information with a high degree of reliability from the beginning of the vehicle tracking stage.
Therefore, RSU puts more weight on the new measurement than the predicted state vector in order to obtain better estimates with less uncertainty.
Moreover, Figs. \ref{fig:sim_result_00} {verify} that the proposed  algorithm outperforms the discrete Fourier transform (DFT) based beamforming solutions in \cite{Ref_Sha19} because it can adjust the beam width of the beamformer according to the state estimation performance as well as reduce the uncertainty in the state prediction process based on the acceleration estimation algorithm.

We evaluate the beamforming gain and system throughput of the proposed downlink beamforming algorithm based on the normalized beamforming gain, defined as $\Gamma \doteq \mathrm{E}\big[ {| \bh^{H} \bc|^2}/{\|\bh \|_2^2} \big]$ and the data-rate, defined as $\mathrm{R} \doteq \mathrm{E}\big[ \log_{2}(1+\rho_{\ell}|\bh^{H} \bc|^2 )\big]$.
The initial state vector is set to ${\bt}_{0}=[-50~\textrm{m}, 8.5~\textrm{m}, 60~\textrm{km/h}]^T$.
The normalized beamforming gain and the data-rate of the proposed vehicle tracking algorithm are compared with those of  the vehicle tracking algorithm in \cite{Ref_Sha19} and those of the feedback-assisted state tracking approach.
Communication systems relying on the DFT codebook are also considered in order to evaluate the beamforming performance of the proposed codebook.
The transmit beamformer in DFT scheme $1$ is chosen based on the estimated state vector $\tilde{\bt} _{\tau+\frac{\Omega}{2}}=\bA^{\frac{\Omega}{2}}\hat{\bt}_{\tau}$ at time $\tau+\frac{\Omega}{2}$.
Furthermore, the beamformer in DFT scheme $2$ is chosen based on the state vector $\hat{\bt}_{\tau}$, which is estimated at time $\tau$.

We evaluate the beamforming gain and data-rate of the beamforming system with a single-resolution codebook  including $Q=M$ codewords.
Figs. \ref{fig:sim_result_01} show that the proposed beamforming system provides higher performance  than the DFT-based beamforming systems because the codewords are designed by considering the road structure.
Furthermore, we consider a beamforming system using {a} multi-resolution codebook consisting of a low-resolution codebook with $Q_1=\frac{M}{2}$ codewords and a high-resolution codebook with $Q_2=M$ codewords.
As verified in Figs. \ref{fig:sim_result_02}, a beamformer with a narrow beam width is chosen to maximize the beamforming gain if the system determines that the state estimation performance is good.
If the state estimation performance is  poor, then a beamformer with a wide beam width is chosen to enhance the beam alignment performance.
On the other hand, the DFT-based system  cannot respond to change in the state estimation performance because it only considers the beamformer with a single type of beam width.
It is also observed that the performance gap between the proposed and DFT codebooks widens as the beamformer update period  increases.

\section{Conclusion}
\label{sec:con}
In this paper, we developed a vehicle tracking algorithm that is suitable for seamless V2I communications in fast time-varying channels.
Based on the EKF algorithm, we proposed a vehicle tracking algorithm  to track the movement of vehicles using limited channel feedback resources.
To realize reliable V2I communications with the high beamforming gain, a beamforming codebook was designed by considering the road structure between the vehicle and RSU.
Furthermore, we developed a beamformer selection algorithm to obtain a robust beamformer by predicting the future motion states of the vehicles.
Simulation results demonstrated that the proposed vehicle tracking  and downlink beamforming algorithms show the improved beamforming gain and data-rate performance although they utilize fewer network resources than the feedback-assisted state tracking approach.

\bibliographystyle{IEEEtran}
\bibliography{ref}

\end{document}